\documentclass[twocolumn]{aastex62}

\usepackage{amsmath}
\usepackage{graphbox,graphicx}
\usepackage{times}
\usepackage{color}
\usepackage{hyperref}
\hypersetup{
  colorlinks=true,       % false: boxed links; true: colored links
  linkcolor=blue,        % color of internal links
  citecolor=cyan,        % color of links to bibliography
}
\newcommand{\pfrac}[2]{\left(\frac{#1}{#2}\right)}
\newcommand{\bfrac}[2]{\left[\frac{#1}{#2}\right]}

\newcommand{\cf}{cf.,~}
\newcommand{\ie}{i.e.,~}
\newcommand{\eg}{e.g.,~}

\shorttitle{When did the remnant of GW170817 collapse to a BH?}
\shortauthors{Gill, Nathanail and Rezzolla}
\begin{document}

\title{WHEN DID THE REMNANT OF GW170817 COLLAPSE TO A BLACK HOLE?}

\author{Ramandeep Gill}
\affil{Institut f\"{u}r Theoretische Physik, Max-von-Laue-Strasse 1, D-60438 Frankfurt, Germany}
\affil{Department of Natural Sciences, The Open University of Israel, 1 University Road, POB 808, 
Raanana, 4353701, Israel}

\author{Antonios Nathanail}
\affil{Institut f\"{u}r Theoretische Physik, Max-von-Laue-Strasse 1, D-60438 Frankfurt, Germany}

\author{Luciano Rezzolla}
\affil{Institut f\"{u}r Theoretische Physik, Max-von-Laue-Strasse 1, D-60438 Frankfurt, Germany}

\begin{abstract}
The main hard pulse of prompt gamma-ray emission in GRB$\,$170817A had a
duration of $\sim0.5\,{\rm s}$ and its onset was delayed with respect to
the gravitational-wave chirp signal by $t_{\rm del} \approx 1.74\,{\rm
  s}$. Detailed follow-up of the subsequent broadband kilonova emission
revealed a two-component ejecta -- a lanthanide-poor ejecta with mass
$M_{\rm ej,blue}\approx0.025\,M_\odot$ that powered the early but rapidly
fading blue emission and a lanthanide-rich ejecta with mass $M_{\rm
  ej,red}\approx 0.04\,M_\odot$ that powered the longer lasting redder
emission. Both the prompt gamma-ray onset delay and the existence of the
blue ejecta with modest electron fraction, $0.2\lesssim Y_e\lesssim0.3$,
can be explained if the collapse to a black hole was delayed by the
formation of a hypermassive neutron star (HMNS). Here, we determine the
survival time of the merger remnant by combining two different
constraints, namely, the time needed to produce the requisite blue-ejecta
mass and that necessary for the relativistic jet to bore its way out of
the expanding ejecta. In this way, we determine that the remnant of
GW170817 must have collapsed to a black hole after $t_{\rm
  coll}=0.98_{-0.26}^{+0.31}\,{\rm s}$. We also discuss how future
detections and the delays between the gravitational and electromagnetic
emissions can be used to constrain the properties of the merged object.
\end{abstract}


%%%%%%%%%%%%%%%%%%%%%%%%%%%%%%%%%%%%%%%%%%%%%%%%%%

%%%%%%%%%%%%%%%%% BODY OF PAPER %%%%%%%%%%%%%%%%%%

%%%%%%%%% INTRODUCTION %%%%%%%%%%%%%%%%%%%%%%%%%%%%%%%%%%%%%%%%%%%%%%%%%%%%
\section{Introduction}

For more than 25 years, binary neutron-star (BNS) mergers have been
hypothesized as the progenitors of short-hard gamma-ray bursts (GRBs;
\citealt{Eichler+89,Narayan+92}, also see
\citealt{Baiotti2016,Paschalidis2016} and \citealt{Nathanail2018b} for
reviews). The neutron-rich dynamical ejecta in such mergers was also
predicted to be the ideal sites for r-process nucleosynthesis
\citep{Lattimer-Schramm-76}, the radioactive decay of which would power
isotropic and bright kilonova emission (\eg \citealt{Li-Paczynski-98},
also see \citealt{Fernandez-Metzger-16} for reviews). Both of these
predictions and the detection of gravitational waves (GWs) from a BNS
merger were confirmed in the detection of GWs from GW170817
\citep{Abbott+17a} and the almost coincident prompt gamma-ray emission
from the short GRB~170817A \citep{Abbott+17b}. In less than $11\,$ hours,
the optical counterpart was localized \citep{Coulter+17} in the nearby
($\simeq40\,$Mpc) elliptical galaxy NGC 4993 and a large-scale
observational campaign was launched to obtain the most detailed
observations in the UV-optical-NIR energy bands of the kilonova emission
\citep{Abbott+17c}.

The total mass of the merged object, determined by restricting the spins
of the two coalescing neutron stars to be between that inferred from
observed BNSs, is $M_{\rm tot} = 2.74_{-0.01}^{+0.04}M_\odot$
\citep{Abbott+17a}. This is significantly larger than the highest value
of masses measured for neutron stars, which are $1.97\pm0.04M_\odot$
\citep{Demorest+10} and $2.01\pm0.04M_\odot$
\citep{Antoniadis2013}. Furthermore, $M_{\rm tot}$ well exceeds the
maximum neutron-star mass predicted by most of the popular
equation-of-states (EOSs). Therefore, it is expected that the merger
remnant, after having ejected a (small) fraction of the total rest mass,
will ultimately collapse to a black hole (BH).

Whether this occurs promptly or not is an important question. Prompt
collapse to BH formation in a BNS merger can be halted for a short time
due to the formation of an HMNS that is supported against gravitational
collapse by differential rotation, which allows for equilibrium solutions
with masses that are substantially larger \citep{Baumgarte+00, Weih2017}
than those allowed by uniform rotation \citep{Breu2016}. In fact, using
full GR simulations of both equal \citep[\eg][]{Shibata06a, Baiotti08}
and unequal-mass binaries \citep[\eg][]{Rezzolla:2010,Hotokezaka2011},
and for a number of EOSs, it was shown that if $M_{\rm
  tot}\lesssim2.7M_\odot$, then the remnant could be a long-lived HMNS
with a collapse time $t_{\rm coll}\gtrsim10\,{\rm ms}$.

The fact that the remnant does not collapse promptly to a BH suggests
that in such BNS mergers the arrival time of the GW and the prompt
gamma-ray emission will not be coincident and the GRB onset will be
delayed by at least the collapse time, such that $t_{\rm del}>t_{\rm
  coll}$. Of course, in this line of reasoning we are assuming that the
relativistic jet is indeed launched by the BH; this is our working
assumption hereafter and is supported by numerous numerical-relativity
simulations that have shown that magnetically confined jet structures may
form once a torus is present around the BH \citep{Rezzolla:2011,
  Paschalidis2014, Dionysopoulou2015, Ruiz2016, Kawamura2016}.

With the exception of the case of a prompt collapse, determining the
collapse time of the remnant for a binary of given total mass is
extremely challenging, as it depends on the unknown EOS, the rotational
profile of the HMNS, and the numerous dissipation processes (radiative
and not) that operate to distribute angular momentum and bring the star
to uniform rotation. Moreover, self-consistent numerical-relativity
simulations -- either in pure hydrodynamics or in magnetohydrodynamics
(MHD) -- are not particularly accurate after the merger because of the
appearance of strong shocks, which inevitably reduce the convergence
order to first \citep{Rezzolla_book:2013}. As a result, most of the
three-dimensional numerical-relativity simulations are limited to
post-merger timescales of $t\sim10\,{\rm ms}-100\,{\rm ms}$
\citep{Baiotti2016,Paschalidis2016} and longer integration times become
computationally unfeasible; such a timescale is also the one that has
been explored in practice to assess the stability of the HMNS. If the
actual collapse time from numerical simulations is rather uncertain (see
\citealt{Radice2018} for a systematic discussion), the upper limit for
the lifetime of the remnant is on somewhat firmer grounds and has been
estimated to be of $\sim 10^4\,{\rm s}$ \citep{Ravi2014}. While this
upper limit is quite generous, it has been deduced when considering a
variety of possible scenarios for the evolution of the remnant.

Overall, numerical-relativity simulations agree in finding that after the
merger, a double-core structure forms first, which then quickly relaxes
to the the newly formed HMNS that has a highly non-axisymmetric bar-like
structure that rotates very rapidly. This generates a substantial
time-varying quadrupole moment and hence a strong emission of GWs over
the next several tens of milliseconds that, alone, could cause the
remnant to collapse. The law of differential rotation has very specific
properties, with the angular-velocity profile being characterised by a
slowly rotating core and an envelope that rotates at quasi-Keplerian
frequencies \citep{Shibata06a, Kastaun2014}; more importantly, this
behaviour has been found to be essentially universal
\citep{Hanauske2017}, \ie only weakly dependent on the EOS.

Differential rotation in the HMNS is however expected to be damped on
longer timescales, \ie $\gtrsim100\,{\rm ms}$, either by direct coupling
with seed magnetic fields, or via other dissipative effects. In the first
case, magnetic fields can be amplified dynamically after merger because
of a number of instabilities that are expected to develop, with the
Kelvin-Helmholtz \citep{Price06} and the magnetorotational (MRI;
\citealt{Siegel2013}) instabilities being those that give the fastest
exponential growths. Despite some computationally very expensive attempts
having been made to measure with confidence the magnetic-field strength
in the HMNS \citep{Kiuchi2015}, the latter is still quite uncertain;
overall an amplification of at least three orders of magnitude is
expected, thus bringing the initial magnetic fields from their canonical
value of $B\lesssim10^{12}\,{\rm G}$ to bulk values of
$B\gtrsim10^{15}\,{\rm G}$. In the second case, viscous effects, either
``direct'', \ie in terms of pure shear and bulk viscosity of nuclear
matter \citep{Alford2017}, or ``effective'', \ie in terms of the
development of the MRI or neutrino transport, could significantly affect
the distribution of angular momentum in the HMNS \citep{Duez2004b,
  Shibata2017b, Radice2017}.

We here use the properties of the kilonova emission and the delay time
between the GW chirp signal and prompt gamma-ray emission onset to
constrain the collapse time of the remnant. Earlier works have either
constrained the remnant collapse time to a broad range or they have
provided only approximate estimates. For example, using a constant mean
mass-ejection rate \citet{Metzger2018} have shown that a model involving
a rapidly spinning HMNS with an ordered surface magnetic-field strength
of $\approx 10^{14}\,{\rm G}$ and with a lifetime $t_{\rm
  coll}\sim0.1-1\,{\rm s}$ can simultaneously explain the velocity, total
mass, and electron fraction of the blue-ejecta mass. On the other hand,
\citet{Granot+17} have argued that a lifetime $t_{\rm coll}\sim1\,{\rm
  s}$ is favored over prompt collapse as it would naturally explain the
delay time of $t_{\rm del}\approx1.74\,{\rm s}$, given that the
relativistic jet launched after collapse would then have to clear
$\sim1\,{\rm s}$ worth of neutrino-driven wind and dynamical ejecta.

In this work, we present a comprehensive account of the different
mass-ejection channels and provide analytical mass-ejection rates derived
from the results of several numerical simulations from the literature.
In particular, based on the time-dependent rates, we argue that in order
to produce the requisite mass of $M_{\rm ej,blue} \approx 0.025\,M_\odot$
in the lanthanide-poor ejecta that gave rise to the rapidly fading bluer
emission in the UV and optical at early times, the collapse time of the
remnant cannot be much smaller than $t_{\rm coll}\simeq1\,{\rm s}$. More
importantly, we arrive at a similar conclusion from an independent line
of argument, where we model the dynamical evolution of the relativistic
jet launched after the remnant collapses and bores its way out of the
dynamical ejecta. In particular, we show that the observed delay time of
$t_{\rm del}\approx 1.74\,{\rm s}$ between the GW and gamma-ray emission
can be satisfied if the remnant collapsed after about one second.

The rest of the paper is structured as follows. In
Sec.~\ref{sec:prompt-delayed-collapse} we discuss the different merger
remnants obtained in BNS mergers and outline the basic properties that
determine the merger outcome -- prompt or delayed collapse.  A detailed
summary of the different mass-ejection channels along with simple
analytic expressions describing the mass-ejection rates, as well as the
total ejected masses as a function of time are presented in
Sec.~\ref{sec:mass-ejection}. The collapse time of the remnant obtained
from the condition of producing the requisite amount of blue ejecta mass
is calculated in Sec.~\ref{sec:tcoll-ejecta}, while in
Sec.~\ref{sec:BO-time} we present the semi-analytic theory that describes
the dynamical evolution of the jet-cocoon system and calculate the
jet-breakout time. These results are then used, along with the mass
ejection rates, to constrain in Sec. \ref{sec:constraint-time} the
collapse time of the remnant in GW170817. Finally, we discuss the
implications of our results in Sec.~\ref{sec:discussion} and provide our
conclusions in Sec.~\ref{sec:conclusions}.

%%%%% PROMPT COLLAPSE VS DELAYED COLLAPSE %%%%%%%%%%%%%%%%%%%%%%%%%%%%%%%%%
\section{Prompt collapse vs delayed collapse}
\label{sec:prompt-delayed-collapse}

In what follows we review and summarize the two different lines of
argument, stemming either from considerations on the maximum mass of
self-gravitating equilibrium configurations
(Sec.~\ref{sec:Mtot_evidence}), or from the observations of the kilonova
emission (Sec.~\ref{sec:kn_evidence}), that support the idea that the
remnant in GW170817 collapsed to a BH sometime after the merger
\citep[see, e.g.,][]{Margalit2017,Shibata2017c,Rezzolla2017,Ruiz2017}.

We start by recalling that there are four possible
outcomes for the remnant of a BNS merger \citep[\eg][]{Baiotti2016}:
\textit{(i)} prompt collapse to a BH, \textit{(ii)} an HMNS that is
supported by differential rotation and thermal pressure, \textit{(iii)} a
supramassive neutron star (SMNS) that is supported by uniform rotation,
and \textit{(iv)} a stable neutron star\footnote{Of course, both an HMNS
  and SMNS will eventually collapse to a BH, although on much longer (and
  different) timescales. Strictly speaking, therefore, there are only two
  asymptotic outcomes of a BNS merger: a stable NS or a black
  hole.}. Each outcome depends on the total mass of the system and the
underlying EOS. While the former can be obtained with higher certainty by
measuring the masses of each of the neutron stars from the GW detection,
the relatively larger uncertainty in the latter allows various
possibilities for the remnant. However, we here assume that for events
that are associated with short GRBs, the remnant must collapse to a
BH. This condition is necessitated by the fact that a BH is most likely
needed \citep[\eg][]{Rezzolla:2011} to power the relativistic outflow,
which then produces a short GRB due to internal dissipation in the
jet. This would naturally invalidate the existence of any kind of a
long-lived stable neutron star. Nevertheless, the possibility that an
SMNS with magnetar-like surface dipole magnetic fields (\ie
$B_s\gtrsim10^{14}\,{\rm G}$) powers the GRB \citep{Zhang2001,
  Gao-Fan-06,Metzger2008} before its ultimate collapse to a BH cannot be
excluded. Indeed, the existence of such a proto-magnetar could help
explain some puzzling long-term and sustained X-ray emission associated
with the afterglows of a number of short GRBs \citep{Rezzolla2014b,
  Ciolfi2014}, for which the delay time is much larger (\ie $\sim
10^2\,{\rm s}$).

\subsection{Evidence for delayed collapse: maximum mass}
\label{sec:Mtot_evidence}

Determining the threshold mass to prompt collapse is possible via
numerical simulations and this has been exploited to set lower-bound
constraints on the radii of neutron-star models \citep{Bauswein2013,
  Bauswein2017b, Koeppel2019}. In its lowest-order approximation, the
threshold mass, that is the minimum mass above which a self-gravitating
system (\eg the merger remnant) will undergo prompt collapse, can be
expressed in terms of the maximum mass for a nonrotating configuration,
$M_{_{\rm TOV}}$ as ${M_{\rm th}}/{M_{_{\rm TOV}}} \approx 1.415$, where
this expression has an uncertainty of $\Delta M_{\rm th}=0.05\,M_{\odot}$
\citep{Koeppel2019}. Using this expression and taking a conservative
value of $M_{_{\rm TOV}} \simeq 2\,M_{\odot}$, we readily derive that
${M_{\rm th}} \simeq 2.82\,M_{\odot}$, and hence above the value measured
for $M_{\rm tot}$ by \citet{Abbott+17a}\footnote{More detailed analyses,
  carried out by a number of groups and following very different
  approaches have now converged to the conclusion that the observational
  evidence coming from GW170817 hints to a maximum mass for a nonrotating
  neutron star in the range $2.01^{+0.04}_{-0.04}\leq M_{_{\rm
      TOV}}/M_{\odot}\lesssim 2.16^{+0.17}_{-0.15}$ \citep{Margalit2017,
    Shibata2017c, Rezzolla2017, Ruiz2017}, where the lower limit in this
  range comes from pulsar observations \citep{Antoniadis2013}.}. Stated
differently, even taking a conservative point of view, it is unlikely
that the remnant of GW170817 collapsed to a BH \emph{promptly}.

At the same time, it is also reasonable to expect that the remnant did
collapse to a BH \emph{at some point} after the merger, simply based on
the estimates about the maximum mass that can be supported by a
self-gravitating fluid configuration. We recall that \citet{Breu2016}
have shown that a quasi-universal relation (\ie a relation that is
essentially independent of the EOS) exists between the maximum mass that
a neutron star can support via uniform rotation, $M_{\rm max}$, and the
corresponding mass for a nonrotating configuration. Exploring a very
large number of EOSs, the maximum mass was found to be
\begin{equation}
M_{\rm max}=\left(1.20 \pm 0.02 \right)M_{_{\rm TOV}} \simeq
2.40\,M_{\odot} < M_{\rm tot}\,,
\end{equation}
again considering that $M_{_{\rm TOV}} \simeq 2\,M_{\odot}$. Similar
considerations can be also made for configurations that are
differentially rotating, once a precise form for the law of differential
rotation has been chosen. In particular, when considering the most common
law of differential rotation, \ie the $j$-constant law, \citet{Weih2017}
found that a quasi-universal relation between the maximum mass of a
differentially rotating neutron star and the maximum nonrotating mass,
\ie
\begin{equation}
  M_{\rm max, dr} \simeq \left(1.54 \pm 0.05\right)
  M_{_{\rm TOV}} \simeq 3.08\,M_{\odot} > M_{\rm tot}\,,
\end{equation}
once again after taking $M_{_{\rm TOV}} \simeq 2\,M_{\odot}$. In summary,
these two results combined suggest that the remnant of GW170817 is likely
to have collapsed eventually to a BH, but also that this must have
happened when the remnant had lost a certain amount of differential
rotation.

The timescale over which differential rotation is lost depends on the
intervening process. In the presence of magnetic fields, differential
rotation in an HMNS is damped by magnetic braking on the Alfv\'{e}n
timescale \citep{Shapiro-00}
\begin{align}
  \label{eq:alfven}
  t_A & = \frac{R_{_{\rm NS}}}{v_A} \nonumber \\
  & \approx 120\pfrac{B_{\rm int}}{10^{15}\,{\rm G}}^{-1}\pfrac{R_{_{\rm NS}}}{12\,{\rm km}}^{-1/2}
  \pfrac{M}{2.7\,M_\odot}^{1/2}~{\rm ms}\,,
\end{align}
where $v_A := B_{\rm int}/\sqrt{4\pi\rho}$ is the Alfv\'{e}n speed in the
neutron-star interior, $\rho$ is its mean rest-mass density, and $B_{\rm
  int}$ is the internal magnetic-field strength. However, differential
rotation in the inner regions of the HMNS can be removed on a much
shorter timescale and of the order of a few tens of dynamical timescales,
\ie $10-15\,{\rm ms}$ \citep{Kastaun2014, Hanauske2017}. Once
differential rotation is lost, and assuming the remnant has not yet
collapsed to a BH, the HMNS will effectively become an SMNS and angular
momentum will be lost over a much longer timescale since the GW is
essentially quenched. This spin-down is likely to be due to
magnetic-dipole radiation, such that the spin-down timescale will be
\begin{align}
  \label{eq:t_sd}
  t_{\rm sd} &= f^{-1}\, \frac{Ic^3}{2\Omega_0^2R_{_{\rm NS}}^6B_s^2} \nonumber \\
  & \approx \frac{1.1\times10^4}{f}\pfrac{B_s}{10^{14}~{\rm G}}^{-2}
\pfrac{P_0}{1~{\rm ms}}^2\pfrac{R_{_{\rm NS}}}{12~{\rm km}}^{-6}~{\rm s}\,,
\end{align}
where $I \simeq 10^{45}~{\rm g~cm}^2$ is the moment of
inertia\footnote{Note that this estimate for the moment of inertia is
  probably an upper limit as only the inner core of the remnant with a
  radius of $\sim 5\,{\rm km}$ is in uniform rotation
  \citep{Hanauske2016}. Furthermore, part of the rotational kinetic
  energy used to derive Eq.~\eqref{eq:t_sd} is actually not lost to
  dipolar radiation, but goes into the ejected mass.}, $B_s$ is the
initial surface dipole magnetic-field strength, $\Omega_0:=2\pi/P_0$ is
the initial angular spin frequency, and $P_0$ is the initial spin
period. Setting $\theta_B$ as the inclination angle between the magnetic
dipole axis and the rotational axis, the parameter $f$ depends on the
details of the model and is given by $f = \chi\sin^2\theta_B$ in vacuum,
with $\chi$ being a constant ($\chi=2/3$) in Newtonian gravity, while a
function of the stellar compactness in general relativity, \eg $\chi \sim
2-3$ \citep{Rezzolla2001,Rezzolla2001_err}; on the other hand,
$f=1+\sin^2\theta_B$ in force-free models \citep{Spitkovsky-06}.

Since most of the rotational energy of the star is released at the
spin-down time, this timescale has been linked to the plateau phases
observed in the X-ray lightcurves of many short-hard GRBs
\citep[\eg][]{Gompertz+13,Rowlinson+13,Zhang2001,Metzger+11}. In this
scenario, the rotational spin-down of the SMNS would power the late-time
energy injection into the relativistic outflow, followed by the collapse
of the SMNS to a BH that would manifest as a sudden drop in the X-ray
flux \citep[however see, \eg][for an alternative
  explanation]{Rezzolla2014b, Ciolfi2014}.

In the case of GW170817, if the remnant was an SMNS, then the large
rotational energy released in the form of an isotropic MHD wind would
have produced a spin-down luminosity of $L_{\rm sd}\gg10^{42}\,{\rm
  erg~s}^{-1}$ for a surface dipole magnetic field strength of
$B_s\leq10^{15}\,$G. However, the observed bolometric luminosity at
$t\lesssim10\,$days post-merger was $L_{\rm bol}<10^{42}\,{\rm
  erg~s}^{-1}$ and cannot be explained even if the surface dipole field
was as high as $B_s\leq10^{16}\,$G \citep[\eg][]{Margutti+18}. In
addition, such a high rate of energy injection by the spinning down SMNS
would have powered an exceptionally bright afterglow emission that was
not observed. Both of these reasons ruled out the possibility of an SMNS
remnant \citep[\eg][]{Granot+17,Margalit2017,Pooley+18}.

%%%%%% FIGURE %%%%%%%%%%%%%%%%%%%%%%
\begin{figure*}
\centering
\includegraphics[width=0.35\textwidth,align=m]{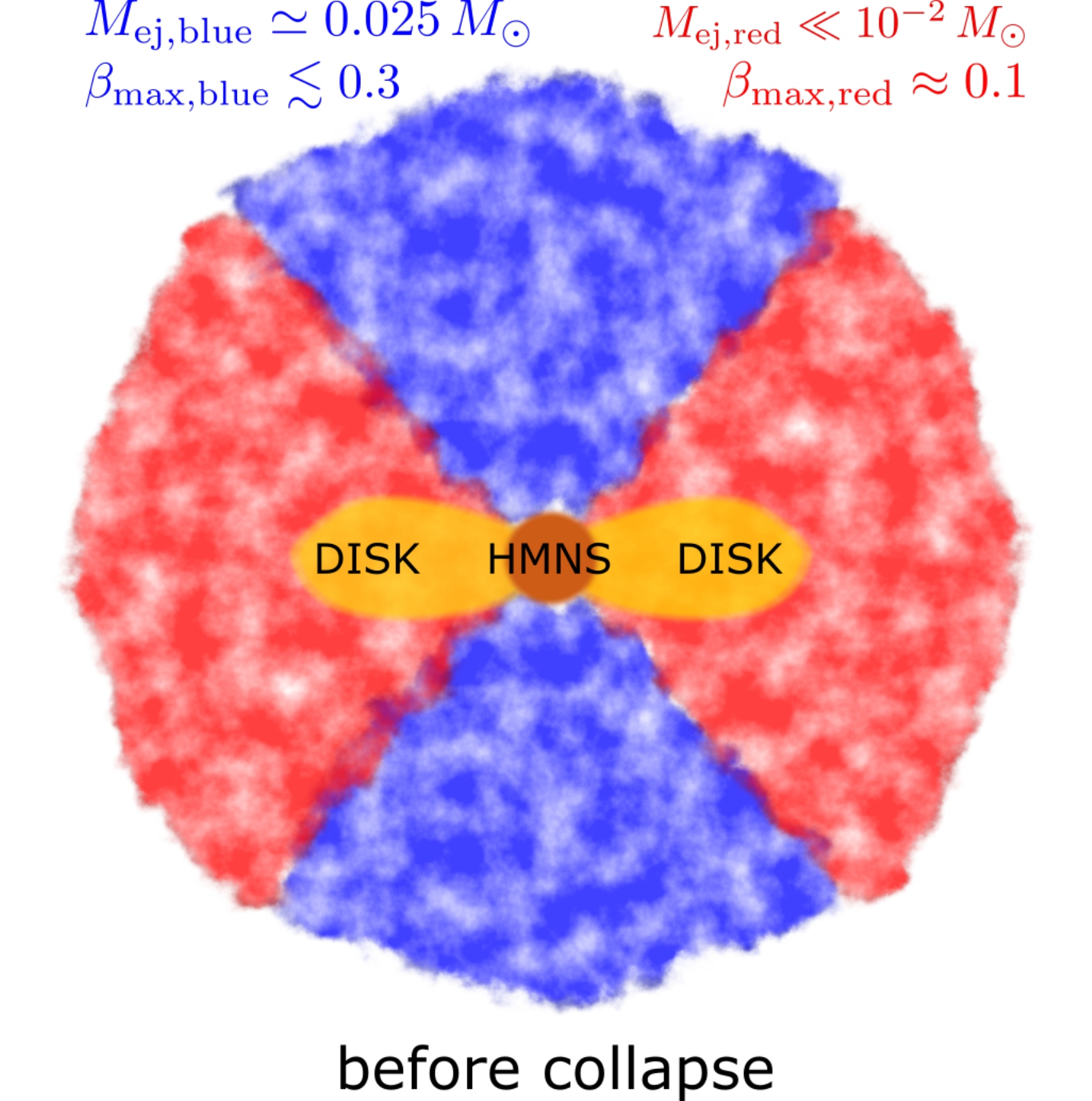}
\hspace{10em}
\includegraphics[width=0.45\textwidth,align=m]{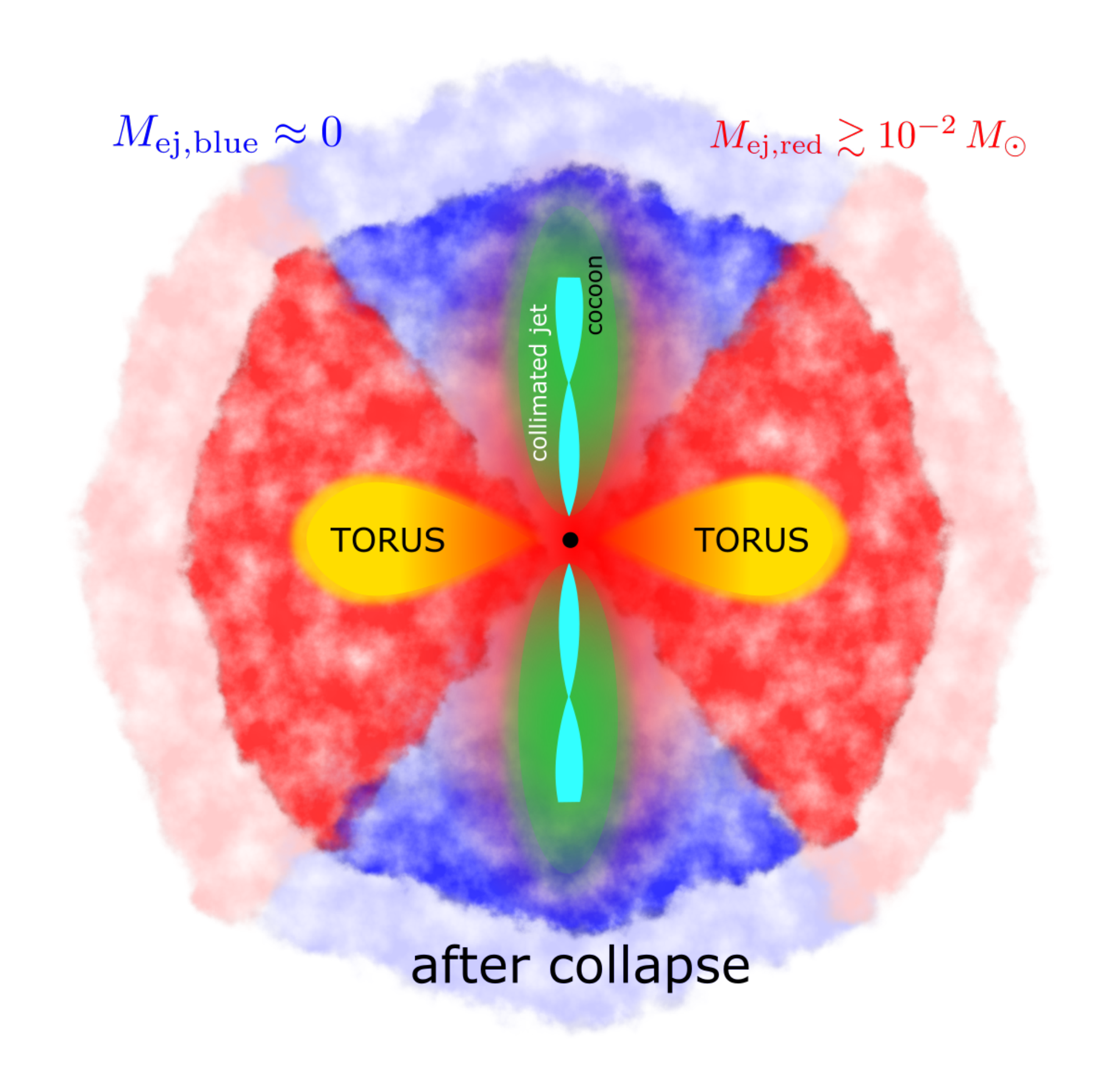}
\caption{Illustration of the mass distribution before and after the
  collapse to a BH of the remnant. Before collapse (left panel), mass is
  ejected dynamically and due to neutrino emission, as well as MHD and
  viscously-driven winds from the HMNS and the ``disk''. This results in
  a two-component ejecta: a fast moving ($\beta_{\rm max,blue}$)
  lanthanide-poor ejecta with mass $M_{\rm ej,blue}$ and with high
  electron fraction ($0.2\lesssim Y_e\lesssim0.3$) that is responsible
  for the early rapidly fading bluer emission, and a relatively slower
  moving ($\beta_{\rm max,red}$) lanthanide-rich ejecta with mass $M_{\rm
    ej,red}$ and with low $Y_e\lesssim0.2$ that leads to the
  longer-lasting red emission. After collapse to a BH (right panel), an
  accretion torus is formed and the BH launches a relativistic jet. Mass
  ejection occurs from the accretion torus due to MHD and
  viscously-driven winds. As the jet traverses the blue ejecta, it
  initially slows down, and the shock-heated jet and ejecta form a cocoon
  that collimates the jet. The jet breaks out when it clears the
  homologously expanding ejecta and produces prompt gamma-ray emission
  due to internal dissipation.}
\label{fig:m_ej_cartoon}
\end{figure*}
%%%%%%%%%%%%%%%%%%%%%%%%%%%%%%%%%%%%

%%%%% EVIDENCE FOR DELAYED COLLAPSE IN GW 170817 / GRB 170817A %%%%%%%%%%%%%%%%%%%%%
\subsection{Evidence for delayed collapse: kilonova emission}
\label{sec:kn_evidence}

We next consider a second and distinct line of argument, based this time
on the observation of the prompt emission and properties of the outflow
that powered the kilonova emission, which again suggest that the collapse
to BH for the remnant in GW170817 was delayed.

We recall that the detection of GW170817 was also rich in electromagnetic
(EM) counterparts. The UV/optical/IR counterparts that followed the GW
were interpreted as the result of heating due to the decay of freshly
synthesized heavy elements. In addition, the onset of the EM emission,
that is, of the prompt gamma-ray emission, as recorded by Fermi/GBM and
INTEGRAL/SPI-ACS, was delayed by $t_{\rm del}=1.74\pm0.05\,{\rm s}$ with
respect to the GW chirp signal that marks the merger of the two neutron
stars \citep{Abbott+17b}. The spectrum of the rapidly fading early UV
emission was well described by a blackbody and the average velocity of
the photosphere was determined to be $\beta := v/c = 0.3 - 0.2$ over the
course of $\sim 0.6-1\,$days after the merger \citep{Arcavi+17,
  Drout2017, Evans+17, Kasen2017, Kasliwal+17, Kilpatrick+17, Pian+17,
  Shappee+17, Smartt+17}. Such high velocities and the trend towards
redder optical and NIR emission that was seen after $\sim2$ days agree
well with predictions from kilonova modeling \citep[\eg][]{Metzger+10}.

However, such observations also require two separate components in the
outflow, one to explain the early UV or \emph{``blue''} kilonova and
another for the \emph{``red''} kilonova that showed a gradual decline in
flux over the initial $\sim2-3$ weeks. The difference between the two
components is that the outflow that powered the blue kilonova was
comprised of lanthanide-poor material, and therefore had a higher
electron fraction, with $Y_{e,{\rm blue}}\sim0.2 - 0.3$, and lower
opacity $\kappa_{\rm blue}\lesssim1\,{\rm cm}^2\,{\rm g}^{-1}$ to
bound-bound electronic transitions. On the other hand, the outflow that
powered the red kilonova was relatively lanthanide-rich with a much lower
$Y_{e,{\rm red}} \lesssim 0.2$ and much higher opacity with $\kappa_{\rm
  red}\sim10\,{\rm cm}^2\,{\rm g}^{-1}$.

The high-opacity lanthanide-rich ejecta is expected to emerge from the
dynamical ejection of material, predominantly due to tidal stripping
\citep[\eg][]{Rezzolla:2010}, as the two neutron stars merge. The ejected
matter in this case lies mostly in the equatorial plane and is confined
to lower latitudes. On the other hand, the lower-opacity lanthanide-poor
ejecta can originate from neutrino-irradiated winds from the accretion
disk \citep[\eg][]{Just2015, Fernandez2015} and/or from an HMNS
\citep[\eg][]{Perego2014}. This is illustrated in the cartoons in
Fig.~\ref{fig:m_ej_cartoon}, where the left panel reports a scheme of the
matter distribution before the collapse of the remnant. In this case,
mass is ejected dynamically, due to shock-heating, neutrino and
MHD-driven winds from the HMNS and the disk. This results in a
two-component ejecta -- a lanthanide-poor with high electron fraction
($0.2\lesssim Y_e\lesssim0.3$) ejecta that powers the early rapidly
fading bluer emission, and a lanthanide-rich with low $Y_e\lesssim0.2$
ejecta that powered the longer lasting redder emission. The increase in
$Y_e$ at higher latitudes is caused by two effects: \textit{(i)} ejection
of only moderately neutron-rich dynamical ejecta in the polar regions
during merger, and \textit{(ii)} by neutrino irradiation of the ejecta
due to neutrinos from the HMNS. After collapse to a BH (right panel of
Fig.~\ref{fig:m_ej_cartoon}), an accretion torus is formed and the BH is
expected to launch a relativistic jet. Mass ejection at this point
primarily occurs from the accretion torus due to MHD and viscously-driven
winds. As the jet traverses the blue ejecta, it slows down initially and
becomes collimated by the cocoon formed by the shock-heated jet material 
and surrounding ejecta. As the jet clears the homologously expanding ejecta, 
it breaks out and produces prompt gamma-ray emission as a result of internal 
dissipation.

Had the remnant of GW170817 suffered a prompt collapse, the newly formed
BH would launch a relativistic jet almost immediately, apart from a
negligible delay on the order of the dynamical time in assembling an
accretion torus. In this scenario, the motion of the jet would be
essentially unimpeded, since most of the ejected mass would be in the
equatorial plane and at high co-latitudes with $\theta > 30^\circ$
(measured from the jet symmetry axis; for an example see Fig.~10 of
\citealt{Bovard2017}, bottom row, third panel from the
left). Furthermore, the jet would cross it in negligible time if any
small amount of circum-merger ejecta were to be present in its
path. Therefore, in the case of prompt collapse, the only time delay
would be the radial-time delay $t_R := R_\gamma/2\Gamma^2c$ for emission
coming from material along our line-of-sight, due to the
less-than-speed-of-light motion of the relativistic flow in reaching the
emission radius $R_\gamma$. However, as discussed in the previous section
and as we will show in Sec.~\ref{sec:BO-time} with more detailed
calculations, $t_R < t_{\rm del}$, thus making the prompt-collapse
scenario problematic for GRB~170817A.

A prompt collapse also creates two additional tensions with the
observations. First, the disk-wind component alone is not able to eject
enough material in order to satisfy the total ejected mass needed for the
blue component. Second, it cannot yield high electron fractions
($Y_e\sim0.3$) at high latitudes that are needed to explain the blue
kilonova. Both of these issues can be settled if the central remnant is
an HMNS. First, hotter surface temperatures lead to a significantly
higher level of neutrino-irradiation, which ultimately raises $Y_e$ to
larger values that can be as high as $Y_e\lesssim0.45$. Second, it was
argued by \citet{Metzger2018} that an enhancement in the mass-ejection
rate and the inferred velocity of $\beta_{\max,{\rm blue}}\sim0.2-0.3$
can be obtained via a strongly magnetized, neutrino-irradiated wind of an
HMNS that possesses magnetar strength surface dipole fields. In this
case, mass ejection is aided by magnetocentrifugal acceleration along
open magnetic field lines, whereas mass ejection due to the thermal
pressure by neutrino-heating alone would yield an insufficient $M_{\rm
  ej,blue}$ and a smaller asymptotic ejecta velocity of $\beta_{\max,\rm
  blue}\sim0.1$. On the other hand, if the collapse was delayed due to
the formation of an HMNS, which survived for $\lesssim1\,{\rm s}$, then
including the breakout time of the relativistic jet from the
quasi-spherically expanding ejecta into the total delay time would
naturally explain the delayed onset of the prompt gamma-ray emission
\citep[\eg][]{Granot+17}.

In the following three sections we will discuss how to set constraints on
the actual time of collapse of GW170817's remnant by modelling either the
ejected mass or EM delay.

%%%%%%% MASS EJECTION %%%%%%%%%%%%%%%%%%%%%%%%%%%%%%%%%%%%%%%%%%%%%%%%%%%%%
\section{Modelling the mass ejection}
\label{sec:mass-ejection}

As mentioned above, the UV/optical/IR observations following GW170817 are
normally explained as the result of heating due to the decay of freshly
synthesized heavy elements contained in the mass ejected during and after
the merger of the binary. The ejected mass was found to be $M_{\rm ej}
\approx 0.02 - 0.06 \, M_{\odot}$, which was confirmed by several
different groups \citep{Chornock2017, Cowperthwaite2017, Drout2017,
  Nicholl2017, Tanaka2017, Tanvir2017, Perego2017, Villar2017,
  Waxman2017, Metzger2018, Kawaguchi2018}. For convenience, we follow
\citet{Kasen2017}, whose kilonova analysis revealed a two-component
ejecta -- a lanthanide-poor ejecta with mass $M_{\rm
  ej,blue}\approx0.025\,M_\odot$ and a lanthanide-rich ejecta with mass
$M_{\rm ej,red}\approx 0.04\,M_\odot$, and use these values to constrain
the lifetime of the remnant and indicate the collapse time.

We recall that BNS mergers are expected to eject mass in a variety of
different processes. In the last few years, thanks to numerical
simulations and semi-analytic modelling, a robust picture has been drawn
on what are the different components of ejected mass and which can be divided
into three main ejection mechanisms:
\begin{itemize}
\item[]\textit{1. matter dynamically ejected} \citep{Rosswog1999,
  Shibata05c, Baiotti08, Kiuchi2010, Rezzolla:2010, Rezzolla:2011,
  Hotokezaka2011, Palenzuela2013, Hotokezaka2013, Sekiguchi2015,
  Palenzuela2015, Dietrich2016, Radice2016, Lehner2016, Sekiguchi2016,
  Foucart2016a, Bovard2017, Dietrich2017, Dietrich2017b, Radice2018a,
  Papenfort2018};

\item[]\textit{2. matter ejected via neutrino-driven winds}
  \citep{Dessart2009, Metzger2008, Metzger09b, Lee2010, Fernandez2013,
    Just2015, Perego2014, Martin2015};

\item[]\textit{3. matter ejected via magnetically-driven winds}
  \citep{Siegel2014, Siegel2017, Ciolfi2017, Fernandez2015,
    Fernandez2018, Fujibayashi2017, Siegel2018, Fujibayashi2017b}.

\end{itemize}

We note that some works also discuss ``viscosity-driven'' ejected
matter \citep{Metzger2008, Goriely2011, Fernandez2013, Fujibayashi2017b,
  Fahlman2018}. These viscous effects, however, are ultimately originated
by magnetic turbulence driven by magnetic instabilities, either in the
HMNS disk or the torus; because of their magnetic origin, we will
hereafter associate them to category \textit{3}.

All of these channels of ejection of matter have been explored in a large
number of works and by several groups employing rather different
numerical techniques and approximations \citep[\eg][]{Just2015,
  Fernandez2015}.
%% While each of these works make
%% contributions to the general picture, a single systematic description of
%% the various processes leading to the ejection of matter, hence a general
%% picture of this process, has not been presented before to the best of our
%% knowledge. For this reason, 
In what follows we discuss the mass-ejection rates for each channel
separately, and provide simple analytic prescriptions that provide simple
and hopefully robust representations of the complex results of the
numerical simulations.

%%%%%% DYNAMICAL EJECTION %%%%%%%%%%%%%%%%%%%%%%%%%%%%%%%%%%%%%%%%%%%%%%%%%%%%%
\subsection{Dynamical mass ejection}

Dynamical mass ejection is a robust feature of BNS merger simulations,
and has been reported by a number of groups under a variety of physical
conditions. The exact amount of mass ejected, however, depends both on
the properties of the binary (mass, mass ratio, EOS, initial spin,
magnetization, etc.) and on the criterion used to perform the
measurement, which could lead even to a $100\%$ difference \citep[see,
  \eg][for a discussion]{Bovard2016}. Furthermore, it is presently
difficult to predict with even decent precision the amount of matter to
be ejected without a fully dynamical simulation as the data does not seem
to be well captured by simple analytic expressions. To counter this
problem, we have exploited the data reported by \citet{Radice2018a}, who
have presented measurements for more than 30 simulations. For the sake of
the arguments made here, we simply need an average of the amount of
matter ejected dynamically, which we deduce from the values for the binaries 
in the mass range $2.7\, M_{\odot} \leq M \leq 2.8\, M_{\odot}$, reported in
Table~2 of \citet{Radice2018a}. In this way, we find a mean value with a
considerable variance, \ie $M_{\rm dyn} \approx (1.5 \pm 1.1) \times
10^{-3} \, M_{\odot}$; although the uncertainty associated to this estimate,
is obviously very large, it also represents the most reasonable one at
present and reflects many of the uncertainties discussed in
\citealt{Bovard2016, Bovard2017}.

\begin{figure*}
    \centering
    \includegraphics[width=0.48\textwidth]{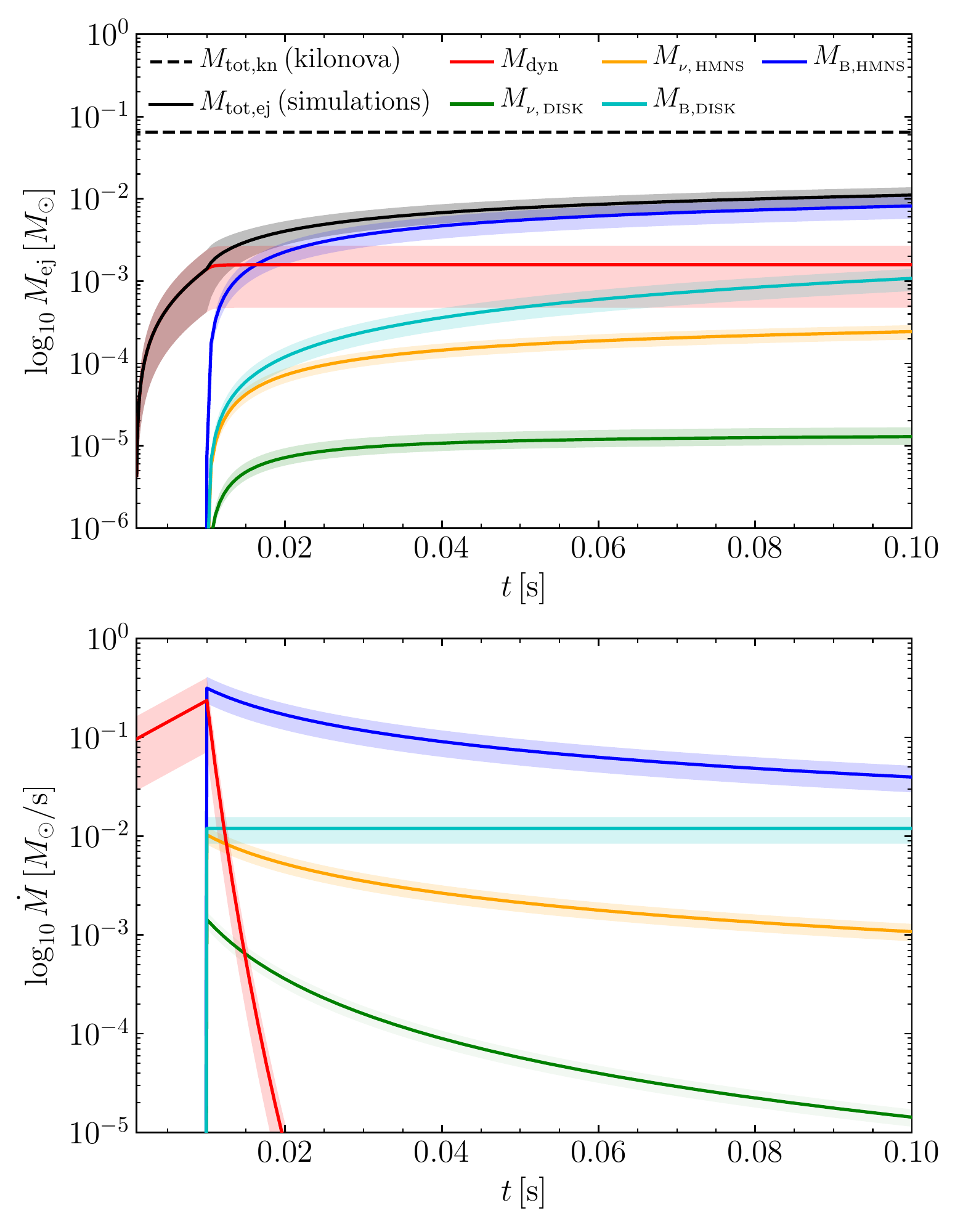}
    \includegraphics[width=0.48\textwidth]{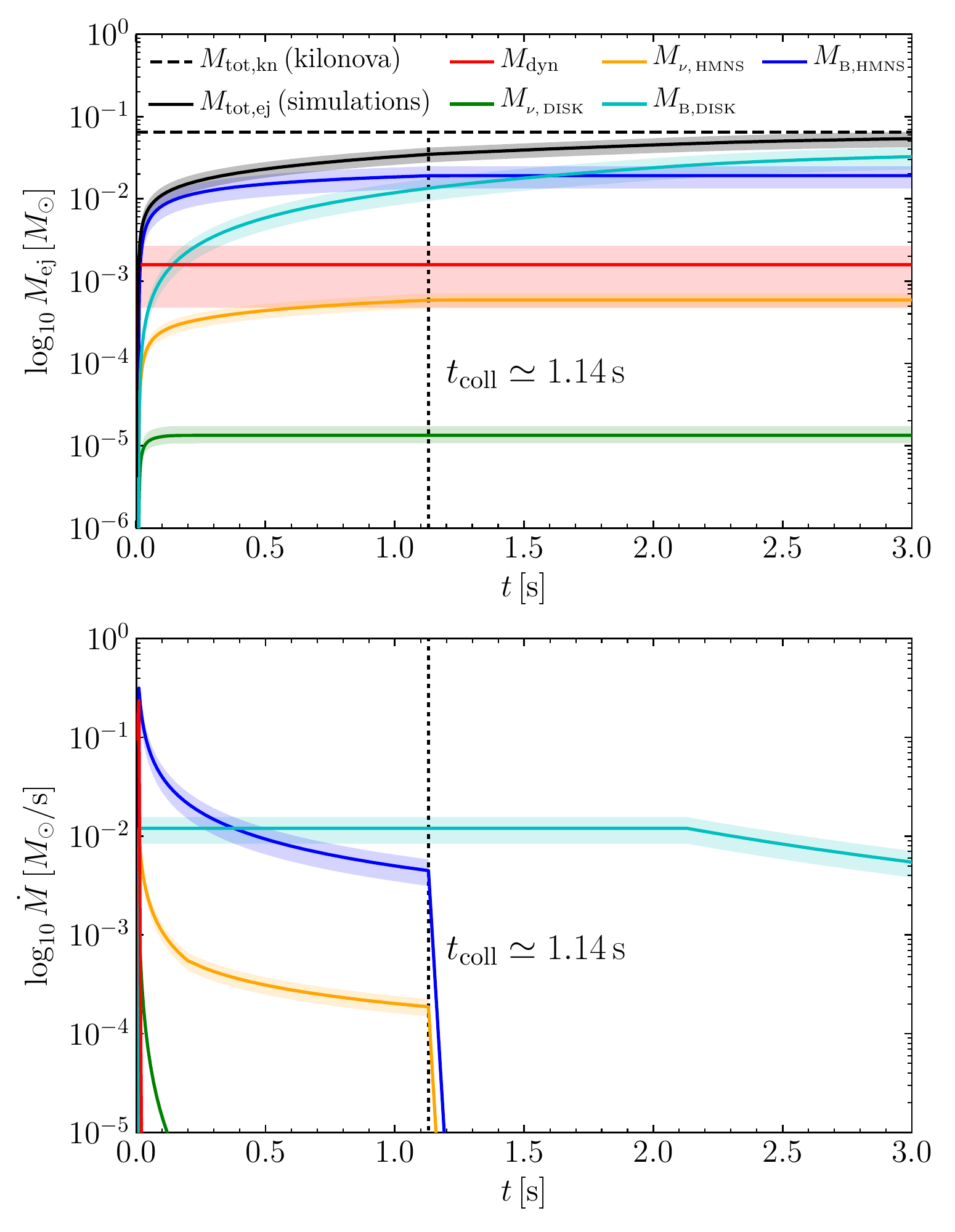}
    \caption{Ejected matter (top plots) and its rate of ejection (bottom
      plots). Different lines and shaded areas refer to the different
      forms in which matter is ejected after the merger and the
      corresponding uncertainty. In particular, we report matter that is
      either ejected dynamically $M_{\rm dyn}$, ejected via MHD-driven
      winds either from the HMNS $M_{_{\rm B, HMNS}}$, or from the disk
      $M_{_{\rm B, DISK}}$, ejected via neutrino-driven winds either from
      the HMNS $M_{_{\nu,\,{\rm HMNS}}}$, or from the disk
      $M_{_{\nu,\,{\rm DISK}}}$. The left panels refer to the first
      $100\,{\rm ms}$, while the right ones to a time interval of
      $3\,{\rm s}$. Shown in black is the sum of the various masses,
      while the dashed horizontal line refers to the limits on the
      ejected mass deduced from the kilonova emission, \ie the sum of the
      blue and red ejecta (see also Fig.~\ref{fig:red_blue}).}
    \label{fig:m_ej}
\end{figure*}

The data reported by \citet{Radice2018a} also allows us to reconstruct a
mass-ejection rate, which we approximate as given by exponential growth
followed by a power-law decay fitted to the data from the numerical
simulations
%
%\begin{equation}
%  \label{eq:mdot_dyn}
%  \dot{M}_{\rm dyn} = \left\{
%\begin{array}{llr}
%  \!\!0.054 \, \exp\left(t/t_{\rm dyn}\right) & M_{\odot}\,{\rm s}^{-1}\,, 
%  &t<t_{\rm dyn}\,, \\
%  \!\!1.4 \times 10^{-31} \, \left({t}/{1\,{\rm s}}\right)^{-15} & M_{\odot}\,{\rm s}^{-1} \,,
%  &t\geq t_{\rm dyn}\,,
%\end{array}
%\right.
%\end{equation}
%
\begin{equation}
  \label{eq:mdot_dyn}
  \frac{\dot{M}_{\rm dyn}}{M_{\odot}\,{\rm s}^{-1}} = \left\{
\begin{array}{lr}
  \!\!0.087 \, \exp\left(t/t_{\rm dyn}\right)\,, & t<t_{\rm dyn}\,, \\
  \!\!2.36 \times 10^{-31} \, \left({t}/{1\,{\rm s}}\right)^{-15}\,, & t\geq t_{\rm dyn}\,,
\end{array}
\right.
\end{equation}
where $t_{\rm dyn}\approx 10 \, {\rm ms}$ is the typical timescale over
which the dynamical ejection has saturated after the merger. In this case, 
we assume a constant fractional uncertainty of 70\%. Here, and
for all of the following expressions for the mass-ejection rate, we set
$t=0$ to be the time of the merger. These dynamical-ejection
contributions are shown in Fig. \ref{fig:m_ej}, where the top panels
report the evolution of the various channels in which mass is ejected
(see also below), while the bottom panels show the corresponding
rates. Shown instead on the left and on the right are the portion of the
evolution on very short timescales after the merger (left panels) and on
timescales comparable with the collapse time (right panels).

%%%%%% FIGURE  %%%%%%%%%%%%%%%%%%%%%%%%%%
%%%%%%%%%%%%%%%%%%%%%%%%%%%%%%%%%%%%%%%%%

%%%%%%% NEUTRINO-DRIVEN WINDS %%%%%%%%%%%%%%%%%%%%%%%%%%%%%%%%%%%%%%%%%%%%%%%%%%%%%%%%%%%%
\subsection{Neutrino-driven winds}

After merger, neutrino-driven winds are expected to be generated from the
remnant and the surrounding disk. Determining the mass-ejection rate from
the various simulations performed to model this process is made difficult
by the fact that only rarely the mass ejected via neutrinos is
reported. Fortunately, however, a close relation exists between the
neutrino-driven mass-ejection rate $\dot{M}_{\nu}$ and the neutrino
luminosity $L_{\nu}$ in its dominant components (\ie electron neutrino
and electron antineutrino) and is given by \citep{Qian1996, Rosswog2002b,
  Dessart2009}
\begin{align}
\dot{M}_{\nu} \approx  5\times 10^{-4} 
\left( \frac{L_{\nu}}{10^{52}\,{\rm erg \, s^{-1}}}\right)^{5/3}
M_{\odot} \, {\rm s}^{-1}\, .
\label{mdot_fromnu}
\end{align}
Since most publications report the measured neutrino luminosity,
expression \eqref{mdot_fromnu} allows us to obtain $\dot{M}_{\nu}$ from
$L_{\nu}$. At the same time, since we want to distinguish the mass
ejection before and after collapse, we will consider separately the
emission coming from the HMNS and that coming from the surrounding disk
(see left panel of the cartoon in Fig.~\ref{fig:m_ej_cartoon}).

In particular, we first consider the neutrino-driven winds coming from
the evolution of the HMNS and to this scope we use the simulations in
\citet{Fujibayashi2017}, where a HMNS with a disk is evolved under the
assumption of axisymmetry, but in a general-relativistic framework and for
a timescale of $400\,{\rm ms}$. In particular, we use the neutrino
luminosity reported in Fig.~3 of \citet{Fujibayashi2017} and fit it with
two power-law segments that change at $200\,{\rm ms}$ after merger, as
measured by the simulations. As a result, we deduce that the
mass-ejection rate associated with the HMNS is 
\begin{equation}
 \frac{\dot{M}_{\nu,{\rm HMNS}}}{M_\odot\, {\rm s^{-1}}} \!\approx\!
  \left\{
  \begin{array}{lr} 
    \!\! 1.13 \times 10^{-4}\, (t/1\,{\rm s})^{-0.98}\ \  &t_{\rm dyn}             < t < 0.2 \, {\rm s}\,, \\
    \!\! 2.02 \times 10^{-4}\, (t/1\,{\rm s})^{-0.62}\ \  &\phantom{t_{\rm dyn} < }  t \geq 0.2 \, {\rm s}\,.
  \end{array}
  \right.
\label{HMNS_nu}
\end{equation}

Three remarks should be made on expression \eqref{HMNS_nu}. First, since
the data reported by \citet{Fujibayashi2017} naturally does not
distinguish the mass-ejection rates from the HMNS and that from the disk,
the values reported in \eqref{HMNS_nu} effectively account also for the
contribution from the disk, $\dot{M}_{\nu,{\rm DISK}}$. As we will see
below, however, the latter is at least one order of magnitude smaller and
with a faster decay; hence, it can be neglected in
\eqref{HMNS_nu}. Second, the rates \eqref{HMNS_nu} match the result of
the simulations up to $t = 0.4 \, {\rm s}$ and are extrapolated for
longer times; this is fine given the steep decay after $t = 0.2 \, {\rm
  s}$ and the fact that this is not a dominant contribution to the
ejected mass. Third, expression \eqref{HMNS_nu} also comes with a
fractional uncertainty, which we estimate to be of the order of $20\%$ on the basis
of the discussion made about the convergence of the results in Sec. 3.5
of \citet{Fujibayashi2017}. Finally, the contribution to the mass
ejection from the HMNS-disk will obviously cease once the remnant
collapses to a BH. We model this sudden shut-off in the mass-ejection
rate via an exponential decay on a dynamical timescale $t_{\rm dyn}
\simeq 10\,{\rm ms}$.

The contribution to the mass ejection coming from the disk around the
HMNS is best approximated by exploiting the results of numerical
simulations that include neutrino-driven winds (and subsequent
nucleosynthesis) coming from a system composed of a BH and an accretion
torus \citep{Fernandez2015b, Just2015, Just2016, Siegel2017,
  Fujibayashi2017, Fernandez2018, Siegel2018}. In particular, we have
made use of the results from \citet{Siegel2017}, which are in good
agreement with those of \citet{Fernandez2018}, and fitted the data in the
upper panel of Fig.~13 of \citet{Siegel2017} with two power-laws joined
at $t=0.12\,{\rm s}$, which then yields
\begin{align}
  \label{Lnu_disk}
  \frac{\dot{M}_{_{\nu,\rm DISK}}}{M_\odot\, {\rm s^{-1}}} \!\approx\!
  \left\{
	\begin{array}{lr} 
  \!\!1.43 \times 10^{-7}\, \left({t}/1\,{\rm s}\right)^{-2.0}\, & t_{\rm dyn}< t < 0.12 \, {\rm s}\,, \\
  \!\!1.22 \times 10^{-11}\left({t}/1\,{\rm s}\right)^{-6.4}\,  &\phantom{t_{\rm dyn}<} t\geq 0.12 \, {\rm s}\,,
\end{array}
\right.
\end{align}
with $\dot{M}_{_{\nu,\rm DISK}}\approx\dot{M}_{_{\nu,\rm tor}}$ and
$\dot{M}_{_{\nu,\rm tor}}$ the mass-ejection rate from a BH-torus system.
Note that $\dot{M}_{_{\nu,\rm DISK}}/\dot{M}_{_{\nu,\rm HMNS}} \ll 1$, so
that the assumption made in \eqref{HMNS_nu} that $\dot{M}_{_{\nu,\rm
    HMNS}} + \dot{M}_{_{\nu,\rm DISK}} \approx \dot{M}_{_{\nu,\rm HMNS}}$
is indeed well justified.

To model the mass ejection at times beyond those simulated by
\cite{Siegel2017}, \ie $400\,{\rm ms}$, we simply extrapolate in time the
second segment of the power-law in \eqref{Lnu_disk}. Also in this case,
we attempt to estimate an error in the mass-ejection rate
\eqref{Lnu_disk} by characterising the numerical noise reported in
Fig.~13 of \citet{Siegel2017}, deducing a lower limit on the fractional 
error of expression \eqref{Lnu_disk} of $\sim20\%$. All of these different
neutrino-driven contributions are shown in Fig. \ref{fig:m_ej}.

%%%%%%%% MHD-VISCOSITY DRIVEN WINDS %%%%%%%%%%%%%%%%%%%%%%%%%%%%%%%%%%%%%%%%%%%%%%%%%%%%%%
\subsection{MHD--viscosity-driven winds}

Magnetically-driven winds from the HMNS are expected to be of great
importance for the overall ejected mass, and especially for the
blue-kilonova component \citep{Metzger2018}. The magnetic energy is only
a small contribution of the total energy budget during the inspiral and
magnetic-induced effects are in fact expected to be negligible sometime
before the merger \citep{Giacomazzo:2009mp}. After the merger, however, a
good fraction of the kinetic energy of the system can be converted into
magnetic energy via magnetic shearing and instabilities. High-resolution
simulations of the remnant have shown that instabilities such as the
Kelvin-Helmholtz or the MRI can develop \citep{Siegel2013}, leading to
magnetic energies as large as $10^{50}\,{\rm erg}$
\citep{Obergaulinger10, Kiuchi2015, Giacomazzo:2014b, Kiuchi2017}.
Simulations have also shown that copious winds are a natural outcome of a
highly shearing magnetized HMNS \citep{Siegel2014}.

Despite being potentially the most important one, mass ejection from
magnetically-driven winds has been explored much less than that from
other channels. It is therefore difficult to quantify the mass-ejection
rate from the few works available in the literature \citep{Siegel2014,
  Dionysopoulou2015, Kiuchi2017, Ciolfi2017}. Here, we make use of the
results of \citet{Ciolfi2017}, who have investigated a long-lived
magnetized HMNS and have shown that while the dynamically ejected mass
does not depend on the magnetic-field strength [see the ejected mass for
  the magnetized and unmagnetized binaries in Fig.~23 of
  \citet{Ciolfi2017}]. On the other hand, when the dynamical ejection of
matter has saturated, the magnetized HMNS does eject mass at a larger
rate than in the unmagnetized case and we have found it to be
well-described by a simple power-law of the type $\dot{M}_{_{\rm B,HMNS}}
\propto t^{-0.8}$. In similar general-relativistic MHD simulations that
will be presented in a distinct work \citep{Most2017b}, the mass-ejection
rate has been found to have a similar but different power-law dependence,
\ie $\dot{M}_{_{\rm B,HMNS}} \propto t^{-1}$. Hereafter, we choose an
average value from the two simulations and set the mass-ejection rate due
to MHD-driven winds from the HMNS (and its surrounding disk) to be
\begin{align}
   \frac{\dot{M}_{_{\rm B, HMNS}}}{M_\odot\, {\rm s^{-1}}}
   \approx 5.18\times 10^{-3} \,\left({t}/1\,{\rm s}\right)^{-0.9} 
   \qquad t_{\rm dyn} \leq t\leq t_{\rm coll}\,.
\label{m_B_HMNS}
\end{align}
As is evident from Fig.~\ref{fig:m_ej}, this mass ejection channel
dominates the mass loss at early times and thus we associate a more
conservative fractional error of $30\%$ for this estimate. At the same
time, because of the very small number of works exploring the effect of
these winds from the HMNS, it is prudent to consider also a different
scenario in which a $30\%$ uncertainty is introduced for the exponent in the
power law; we will discuss the two scenarios separately below.

The viscously/MHD driven winds, first from the HMNS-disk and later from
the BH-torus, are expected to be the dominant channel for mass ejection,
especially after the HMNS collapses \citep{Lippuner2017b,
  Fujibayashi2017b, Fahlman2018}. We model this mass ejection using the
long-term simulations of \citet{Fujibayashi2017b} and deduce that the
rate is impressively constant over a considerably long timescale and can
be approximated as [\cf Fig.~8 of \citet{Fujibayashi2017b}]
\begin{align}
\label{Fuji_visc}
\frac{\dot{M}_{_{\rm B, DISK}}}{M_{\odot}\, {\rm s^{-1}}} \simeq 0.012 \qquad
\quad t_{\rm dyn} \leq t\leq t_{\rm coll}\,,
\end{align}
Note that as long as $t < 0.4\,{\rm s}$, $\dot{M}_{_{\rm B, DISK}} \ll
\dot{M}_{_{\rm B, HMNS}}$, so that its contribution in expression
\eqref{m_B_HMNS} can be safely neglected (see right panels of
Fig.~\ref{fig:m_ej} for a graphical impression). In addition, for
$0.4\,{\rm s}< t <t_{\rm coll}$ the contribution of $\dot{M}_{_{\rm B,
    DISK}}$ to the total ejected mass is very small 
    ($\lesssim 9\times10^{-3}\,M_{\odot}$). Furthermore, since the rates show a rather
large variance, here too we associate a conservative $30\%$ fractional 
error for the estimated rate \eqref{Fuji_visc}.

As for the case of neutrino-driven winds, the contribution of the torus
to the mass-accretion rate due to magnetic winds will be slightly
different from that of the disk around the HMNS, but certainly
comparable. Hence, we assume that also in the case of magnetic-related
rates, $\dot{M}_{_{\rm B,DISK}} \approx \dot{M}_{_{\rm B,tor}}$. The
assumption that the two rates $\dot{M}_{_{\rm B, DISK}}$ and
$\dot{M}_{_{\rm B, tor}}$ are comparable is indeed confirmed by the
recent simulations of \cite{Fernandez2018}, who have performed
three-dimensional general-relativistic MHD simulations of a BH-torus
system for very long times, \ie for timescales of the order of $\sim
9\,{\rm s}$. At the same time, the new simulations also reveal that the
rate \eqref{Fuji_visc} is not always constant and decreases rapidly
$\approx 1\, {\rm s}$ after the start of the simulations (thus,
approximately one second after the collapse of the HMNS to a BH)
following a power law $\propto t^{-2.3}$ \citep{Fernandez2018}. Hence, we
model the magnetically-driven mass-ejection rate from the disk/torus as
\begin{align}
\label{M_B_disk}
\frac{\dot{M}_{_{\rm B, DISK}}}{M_{\odot}\,{\rm s^{-1}}}
\approx \left\{ 
\begin{array}{lr} 
  \!\!0.012                                        &t_{\rm coll} < t < \bar{t} \,,  \\
  \!\!0.012 \, \left( {t}/{\bar{t}} \right)^{-2.3}  &\phantom{t_{\rm coll}<} t \geq \bar{t}\,, \nonumber \\
\end{array}
\right.\\
\end{align}
where $\bar{t}:=t_{{\rm coll}} + 1\,{\rm s}$. Also these magnetic
contributions are shown in the various panels of Fig.~\ref{fig:m_ej}.

\section{Collapse time from the ejecta}
\label{sec:tcoll-ejecta}
\label{subs:kilonova}

The lifetime of the remnant from GW170817 can now be conveniently
constrained by comparing the amount of ejected matter discussed in the
previous sections with the amount of matter that is deduced to have been
ejected on the basis of the \textit{blue}-kilonova emission. We recall
that the latter is expected to be the result of lanthanide-poor material
and thus with a lower opacity \citep{Arcavi+17,Drout2017,Kilpatrick+17,
  Kasen2017, Villar2017}. Producing such material requires high
temperatures and high densities, such that neutrino-irradiation processes
can produce material with high electron fractions (\ie neutron poor),
which, in turn, are responsible for light r-process element production
with intrinsically low opacities. Such large temperatures and densities
can be found in the HMNS, but neither in the disk surrounding it, nor in
the torus around the BH once it is formed. On the contrary, the disk
outflows with low-electron fraction (\ie neutron-rich) material, are
expected to produce heavy r-process elements resulting in higher opacity
\citep{Kasen2013, Tanaka2013, Metzger2014, Perego2014, Lippuner2017b,
  Tanaka2018, Fujibayashi2017b}.

Exploiting the mass-ejection rates reported in the previous sections, we
can estimate the collapse of the HMNS by requiring that the accumulated
ejected mass from the HMNS becomes equal to the blue component from the
kilonova modeling, which has been estimated to be $M_{\rm ej,
  blue}\approx0.025\,M_\odot$ \citep{Kasen2017}. In particular, we assume
that all the mass lost by the HMNS contributes in its entirety to the
mass in the blue component, $M_{\rm ej, blue}$. On the other hand, the
mass ejected dynamically, as well as that coming from the disk and torus,
will contribute to both the blue and the red components. More
specifically, we assume that $1/3$ of the dynamically ejected mass
contributes to the blue component, while $2/3$ of it is channelled into
the red component \citep{Bovard2017, Perego2017, Shibata2017b,
  Radice2018a}. The same percentages are applied also for the matter
ejected from the HMNS-disk and these values are in agreement with the
results of \citep{Fujibayashi2017b, Shibata2017b}. On the other hand,
once the HMNS collapses to a BH, the contributions change as only a
smaller fraction of the matter ejected is able to contribute to the blue
component. Following \cite{Fernandez2018}, we distribute $1/12$ of the
torus-ejected matter to the blue component and the remaining $11/12$ to
the red component. Ideally, these splitting factors ought to be a
function of time and have an associated uncertainty; however, for
simplicity we consider them as constant here and factor-in the
uncertainties in the the ejection rates.

In summary, the mass ejection rates \eqref{eq:mdot_dyn}--\eqref{M_B_disk} and 
their contributions to the total blue- and red-ejecta masses
\begin{equation}
M_{\rm ej, blue/red}(t) = \int\dot M_{\rm ej,blue/red}(t')dt'\,,
\end{equation}
where
\begin{align}
  \label{eq:Mej-blue}
  \dot M_{\rm ej,blue}(t) = & \phantom{+} \eta_{\rm dyn,blue}\,\dot M_{\rm dyn}(t)  \nonumber \\
  & + \eta_{_{\rm HMNS,blue}}[\dot M_{_{\rm B, HMNS}}(t) + \dot M_{_{\rm \nu, HMNS}}(t)] \nonumber \\ 
  & + \eta_{_{\rm DISK,blue}}[\dot M_{_{\rm B, DISK}}(t) + \dot M_{_{\rm \nu, DISK}}(t)]   \,, \\
  \label{eq:Mej-red}
  \dot M_{\rm ej,red}(t) = & \phantom{+} \eta_{\rm dyn,red}\,\dot M_{\rm dyn}(t)  \nonumber \\
  & + \eta_{_{\rm DISK,red}}[\dot M_{_{\rm B, DISK}}(t) + \dot M_{_{\rm \nu, DISK}}(t)]   \,, 
\end{align}
where, for $t \leq t_{\rm coll}$
\begin{align}
& \eta_{_{\rm HMNS,blue}} = 1\,, & \eta_{\rm dyn,blue} = 1/10\,,  && \eta_{_{\rm DISK,blue}} = 1/3 \,, \nonumber\\
& &                            \eta_{\rm dyn,red} = 9/10\,,  && \eta_{_{\rm DISK,red}} = 2/3 \,, \nonumber\\
\end{align}
whereas, for $t > t_{\rm coll}$
\begin{align}
&   \eta_{_{\rm HMNS,blue}} = 0\,, & \eta_{\rm dyn,blue} = 0\,,  &&  \eta_{_{\rm DISK,blue}} = 1/12\,, \nonumber\\
& &                              \eta_{\rm dyn,red} = 0\,,  &&  \eta_{_{\rm DISK,red}} = 11/12\,. \nonumber\\
\end{align}

In this way, we can constrain the lifetime of the remnant by determining
when the ejected mass in the blue component reaches the estimated
observational value, \ie $M_{\rm ej, blue}(t_{\rm coll}) = 0.025
\,M_\odot$. Doing so, we deduce that the collapse time as constrained
from the ejecta and from the condition that $t_{\rm coll}\leq t_{\rm
  del}$ must have been at
%  \footnote{A more conservative lower limit on the
%  collapse time is obtained, with $t_{\rm
%    coll}=1.14^{+0.60}_{-0.84}\ \,{\rm s}\,$, when we consider a $30\%$
%  uncertainty in the power-law exponent in the mass-ejection rate
%  (\ref{m_B_HMNS}) from magnetically driven winds; see also
%  Fig. \ref{fig:tcoll-compare-plt}.}
%
\begin{equation}
  \label{eq:tcoll_ejecta}
  t_{\rm coll}=1.14^{+0.60}_{-0.50}\ \,{\rm s}\,.
\end{equation}
%

%%%%%% FIGURE  %%%%%%%%%%%%%%%%%%%%%%%%%%
\begin{figure}
    \centering
    \includegraphics[width=0.48\textwidth]{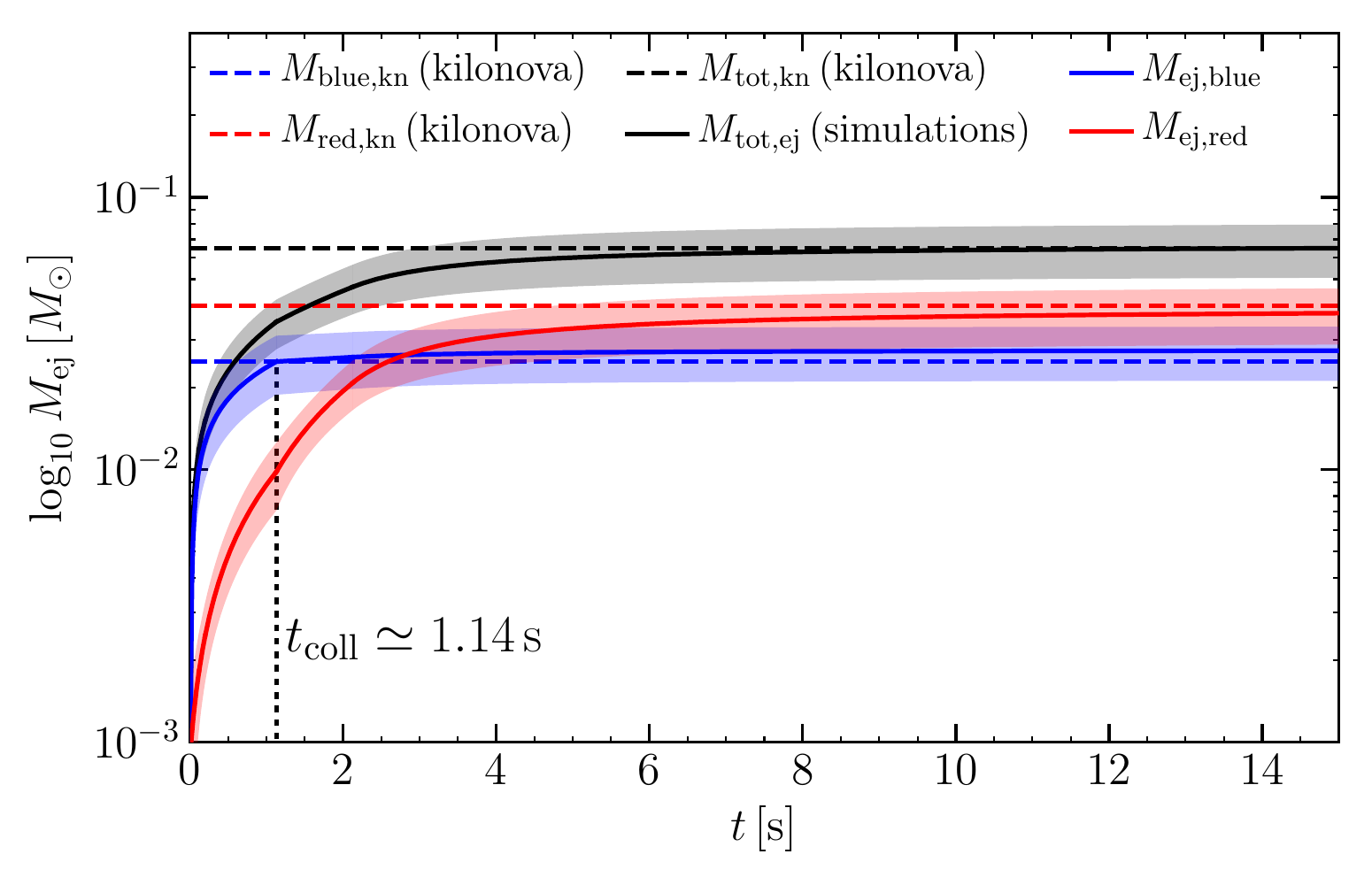}
    \caption{Contributions of the ejected matter to the blue and red
      component of the kilonova emission. Blue/red solid lines show the
      ejected mass in the blue/red components as a result of the various
      channels contributing to these components, while the black solid
      line being the sum of the two. Shaded regions of the corresponding
      colour report the uncertainties in the estimates, while the
      blue/red dashed horizontal lines refer to the limits on the ejected
      mass from the modelling of the kilonova observations.}
    \label{fig:red_blue}
\end{figure}
%%%%%%%%%%%%%%%%%%%%%%%%%%%%%%%%%%%%%%%%%

This is illustrated in Fig.~\ref{fig:red_blue}, where we show the ejected
mass in the blue (blue solid line) and red (red solid line)
components, respectively. Note we set the most likely time of collapse
from the crossing of the ejected blue component with the corresponding
observational estimate (blue dashed line). Note also that the ejected red
component is always less than the blue component and that the latter is
almost three times larger; only around $\approx 2.5\,{\rm s}$ the red and
blue components have been ejected in equal amounts and at $\approx 200\,
{\rm s}$ the red component has reached the nominal value of $M_{\rm
  ej,red}\approx 0.04\,M_{\odot}$ deduced from the observations.  The
very slow increase in the red component is due to the fact that after
$9\, {\rm s}$ the torus has been mostly accreted, thus providing only a
very small contribution to the secularly ejected matter [see
  Eq. \eqref{M_B_disk} and \citet{Fernandez2018}].

The uncertainty in $t_{\rm coll}$ reported in Eq. \eqref{eq:tcoll_ejecta}
was obtained from the corresponding uncertainty in the ejected
blue-component mass. More specifically, the lower limit on $t_{\rm coll}$
was obtained when the upper limit on the blue-component mass crossed the
estimate inferred from kilonova observations. The upper limit, on the
other hand, is limited by the delay time of $t_{\rm del} = 1.74\,$s since
$t_{\rm coll}<t_{\rm del}$. When a constant fractional error $\delta_i$
in the ejected mass $M_i(t)$ from a given mass-ejection channel $i$ was
considered, the final uncertainty in the blue component mass was obtained
by adding in quadrature the uncertainties for the different components
that contributed to the blue-ejecta mass, namely, (\ref{eq:Mej-blue}),
such that $\sigma_{M_{\rm ej,blue}}(t) = \sqrt{\sum_i\sigma_{M_i}^2(t)}$,
where $\sigma_{M_i}(t) :=\delta_i \eta_iM_i(t)$ and $\eta_i$ is their
fractional contribution. Similarly, when considering an uncertainty in
the power-law exponent of the mass ejection rate (\ref{m_B_HMNS}), with
$\dot M\propto t^\alpha$, the uncertainty in the ejected mass from this
channel was calculated as $\sigma_M(t) = \vert(\partial M(t)/\partial
\alpha)\sigma_\alpha\vert$, where $\sigma_\alpha$ is the uncertainty in
the exponent.

We note that a more conservative lower limit on the collapse time is
obtained, with $t_{\rm coll}=1.14^{+0.60}_{-0.84}\ \,{\rm s}\,$, when we
consider a $30\%$ fractional uncertainty in the power-law exponent in the
mass-ejection rate (\ref{m_B_HMNS}) from MHD-driven winds. In this case,
due to a smaller exponent in the power law, the mass ejection rate
increases at times $t<1\,$s as compared to the case with constant
fractional error in the ejected mass. This causes the upper limit on the
blue component mass to cross the blue component mass limit (as shown in
Fig.~\ref{fig:red_blue}) at an earlier time. Hence a smaller lower limit
on the collapse time; see also Fig. \ref{fig:tcoll-compare-plt}.

%%%%%%% JET-BREAKOUT TIME  %%%%%%%%%%%%%%%%%%%%%%%%%%%%%%%%%%%%%%%%%%%%%%%%
\section{Collapse time from Jet Collimation and Breakout}
\label{sec:BO-time}

If in the previous sections we have set constraints on the collapse time
of the remnant in GW170817 from considerations on the ejected mass, in
this section we set a similar constraint following a complementary route
which is instead based on considerations about the time needed by the jet
launched by the BH to break out. Essential in this line of argument is
the velocity of the ejected matter, which, for simplicity, we assume to
be a quasi-isotropic outflow with its front at radius 
$r=R_{\rm ej}(t)=\beta_{\max}ct$ expanding into the interstellar medium 
at a characteristic dimensionless velocity $\beta_{\max} := v_{\max}/c$. 
The mean rest-mass density of the ejecta at any given radius $r<R_{\rm ej}(t)$ can be
parameterized as a power law
\citep[\eg][]{Nagakura+14,Matsumoto-Kimura-18}, such that
\begin{equation}
  \label{eq:rho_ej}
  \rho_{\rm ej}(r<R_{\rm ej},t) =
\frac{(3-k)}{4\pi}\frac{M_{\rm ej,blue}(t)}{R_{\rm ej}^3(t)}\bfrac{r}{R_{\rm ej}(t)}^{-k},~k<3
\end{equation}
where $M_{\rm ej,blue}(t)$ is the mass of the blue ejecta at time
$t$. Our fiducial model assumes $k=2$ for the density profile of the
ejecta along the rotational axis. This profile has been shown to emerge
in the long-term simulations of \citet{Fujibayashi2017b}. Some
modifications to this profile will occur if the remnant is endowed with
strong global magnetic fields, which would alter the mass-ejection rate
due to the MHD winds in the polar regions
\citep[\eg][]{Metzger+07,Siegel2014}.

In our phenomenological modelling, at $t=t_{\rm coll}$, ejection of matter 
from the remnant essentially stops and the central BH launches a relativistic
jet. Naturally, there may be a small delay (of the order of the dynamical
time) between the termination of mass ejection and the launching of the jet as
mass starts to accrete on to the BH; because of its smallness, we here
neglect such a delay and assume that the jet is launched at $t=t_{\rm
  coll}$.

In order for the relativistic jet to reach the radius $r=R_\gamma$, where
it undergoes some internal dissipation and produces the prompt emission,
it must first burrow its way through the intervening ejecta. We assume
that the jet is launched at relativistic speeds with Lorentz factor
$\Gamma_j$ and power per unit solid angle
\begin{equation}
\frac{dL_j}{d\Omega} = r^2\Gamma_j^2h_j\rho_jc^3 \simeq \frac{L_j}{\Delta\Omega_j}\,,
\end{equation}
where $h_j$ and $\rho_j$ are the jet's specific enthalpy and rest-mass
density, respectively. The last equality assumes that the jet power is
approximately uniform and it is confined into a solid angle
$\Delta\Omega_j$.

As the jet expands, its interaction with the surrounding ejecta slows it
down, such that the jet's \emph{head} moves only mildly relativistically
with Lorentz factor $\Gamma_h := (1-\beta_h^2)^{-1/2}<\Gamma_j$, where
$\beta_h:=v_h/c$ is its dimensionless velocity. A double-shock structure
develops at the jet's head (see, \eg Fig.~1 in \citealt{Bromberg+11})
where a forward shock propagates into the cold circum-merger ejecta and a
reverse shocks moves into the cold jet, which slows it down. A contact
discontinuity separates the newly shocked jet material from the shocked
circum-merger ejecta. The velocity of the jet head can be calculated by
balancing the momentum flux density in the frame of the jet's head
\citep[\eg][]{Begelman-Cioffi-89, Matzner-03, Bromberg+11}
\begin{equation}
  \rho_jh_j[\Gamma_j\Gamma_h(\beta_j-\beta_h)]^2 =
  \rho_{\rm ej} h_{\rm ej}[\Gamma_{\rm ej}\Gamma_h(\beta_h-\beta_{\rm ej})]^2\,,
\end{equation}
where $\Gamma_{\rm ej}\simeq1$ is the Lorentz factor of the ejecta and
$h_{\rm ej}$ is its specific enthalpy. Here, the unshocked jet and circum-merger
ejecta are considered cold, and therefore they contribute negligible
pressure. From this relation it is straightforward to obtain the head
velocity
\citep[\eg][]{Matzner-03,Bromberg+11,Murguia-Berthier+14,Murguia-Berthier+17},
\begin{equation}\label{eq:beta-head}
\beta_h = \frac{\beta_j+\tilde L^{-1/2}\beta_{\rm ej}}{1+\tilde L^{-1/2}}\,,
\end{equation}
where the dimensionless jet luminosity, $\tilde{L}$, which determines the
jet dynamics, is given by the ratio of the jet energy density to that of
the surrounding ejecta
\begin{equation}\label{eq:L-tilde}
\tilde L := \frac{\Gamma_j^2h_j\rho_j}{\rho_{\rm ej}} \simeq \frac{L_j}{\Sigma_j\rho_{\rm ej} c^3}\,.
\end{equation}
Here $\Sigma_j := z_h^2\Delta\Omega_j$ describes the area of the working
surface at the head of the jet as it bores its way out of the ejecta,
where $z_h(t) :=\int\beta_hcdt$ is the position of the jet's head. For a
conical (\ie uncollimated) flow, $\Delta\Omega_j = \pi\theta_0^2$ where
$\theta_0\ll1$ is the initial jet half-opening angle. In this case, the
jet expands ballistically, with $\Sigma_j\propto z_h^2$, and the feedback
of the hot cocoon on the jet itself is ignored (see discussion below). In
principle, the jet power can vary over time due to the intermittent
activity of the central BH. For simplicity, however, we here assume that
jet power remains constant in time, so that $L_j$ effectively represents
the time-averaged jet power.

Because of the high-pressure shocked ejecta downstream of the forward
shock, the shocked jet material and the shocked ejecta are able to move
sideways away from the jet's head until the angular size of the shocked
causal region, $\Gamma_h^{-1}$, becomes smaller than the instantaneous
half-opening angle of the jet, \ie $\Gamma_h^{-1}<\theta_j(t)$. Before this
condition is met, a jet-cocoon system is produced, where the pressure in
the hot cocoon collimates the relativistic jet and reduces its
semi-aperture from its initial value $\theta_0$ (see Fig.~1 of
\citealt{Harrison+18}). Here we follow the treatment of
\citet{Bromberg+11}, which applies for a static ambient medium into which
the jet expands, to write the equations that govern the cross-section of
the working surface at the jet's head. Of course, this implies that the
local velocity of the ejecta, which is modeled here as expanding
homologously with dimensionless velocity
\begin{equation}\label{eq:beta-ejecta}
    \beta_{\rm ej}(r<R_{\rm ej},t) = \beta_{\max}\pfrac{r}{R_{\rm ej}(t)}\,,
\end{equation}
must be significantly smaller than that of the jet's head, \ie $\beta_{\rm
  ej} \ll \beta_h$. The assumption of homologous expansion, as compared to a constant 
  velocity wind, is useful here since at a smaller radius the ejecta velocity will also 
  be smaller when the jet is launched. Given that the jet head will initially be 
  sub-relativistic, a slower ejecta velocity helps to preserve the condition of 
  having $\beta_{\rm ej} < \beta_h$ throughout most of the jet propagation inside 
  the expanding ejecta. Under this assumption, the cocoon pressure, assumed
here to be dominated by radiation, depends on the energy injected by the
jet, such that
\begin{equation}\label{eq:cocoon-pressure}
p_c = \frac{1}{3}\frac{E_c}{V_c} = \frac{1}{3}\frac{L_j \int(1-\beta_h)dt}{\pi r_c^2(t)z_h(t)}\,,
\end{equation}
where $r_c(t) := \int\beta_ccdt$ is the cocoon radius expanding with
dimensionless velocity
\begin{equation}\label{eq:beta-cocoon}
    \beta_c := \bfrac{p_c}{\rho_{\rm ej}(z_h)c^2}^{1/2}\,.
\end{equation}

Our model assumes that the pressure in the cocoon is uniform and its
geometry is that of a cylinder. Therefore, under this simplification we
study the dynamics of the jet's head along the jet axis only,
$z_h(t)$. The jet is injected at a radius $z_{\rm inj}$, after which it
starts to be influenced by the cocoon as the pressure in the cocoon
builds up. The location $z_{cj} > z_{\rm inj}$, where the cocoon
collimates the jet, and after which the jet maintains a constant
cylindrical radius, is derived from the condition of pressure
equilibrium, $p_c=p_j$, which yields
\begin{equation}\label{eq:zcj}
    z_{cj} \simeq \bfrac{L_j}{\pi c p_c}^{1/2}\,.
\end{equation}
However, the cocoon starts collimating the jet at an even smaller
distance of $z\approx z_{cj}/2$, at which point the area of the working
surface can be expressed as 
\begin{equation}
  \label{eq:sigma-j}
  \Sigma_j = \pi r_j^2\simeq \frac{1}{4}\pi\hat
  z^2\theta_0^2 \simeq \frac{1}{4}\frac{L_j\theta_0^2}{c p_c}\,,
\end{equation}
where $r_j$ is the cylindrical radius of the jet's head. 

The set of equations presented above,
Eqs.~(\ref{eq:beta-head})--(\ref{eq:sigma-j}), and which have been
introduced mostly by \citet{Bromberg+11}, represent a closed system that
we have solved numerically to follow the evolution of the jet-cocoon
system. Under the simplifying assumption that the velocities of the
different components in the system remain constant, a linearized set of
equations can be obtained that can be integrated analytically. Then, for
a static medium, it can be further shown that the jet interacts strongly
with the surrounding medium and remains collimated for $\tilde
L\lesssim\theta_0^{-4/3}$ \citep{Bromberg+11}. In an expanding medium,
this result does not apply strictly and deviations from it are expected.

As discussed above, these equations are based on several simplifying
assumptions, \eg of uniform pressure in the cocoon and its cylindrical
geometry, or the fact that only the first collimation shock, which is
much stronger than the subsequent ones~\citep{Mizuno2015}, is accounted
for. These assumptions, however, were verified to provide a reasonable
approximation in a comparison with two- and three-dimensional
hydrodynamic simulations~\citep{Harrison+18}.

%%%%%% FIGURE  %%%%%%%%%%%%%%%%%%%%%%%%%%
\begin{figure}
    \centering
    \includegraphics[width=0.48\textwidth]{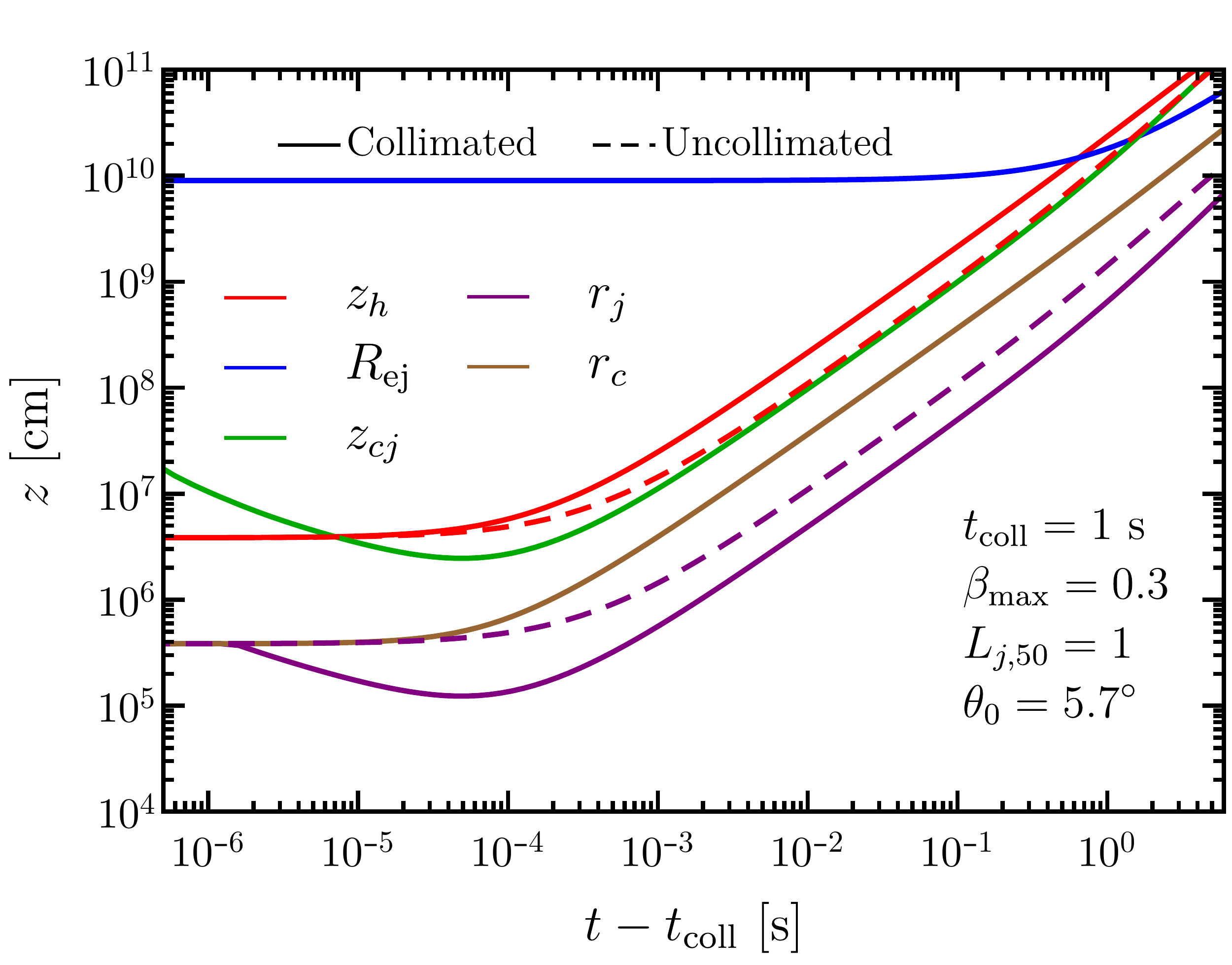}
    \caption{Dynamical evolution of the jet-cocoon system, where the HMNS
      survives for $t_{\rm coll}=1\,{\rm s}$ before collapse to a BH,
      which launches a jet with power $L_j = 10^{50}\,{\rm erg~s}^{-1}$
      and initial half-opening angle $\theta_0=\Gamma_j^{-1}\simeq
      5.7^\circ$.  The jet is injected into the ejecta expanding
      homologously with its front moving at $\beta_{\max}=0.3$, and which
      has a density profile of that of a wind ($k=2$) with total ejected
      mass calculated from the mass-ejection rates shown in
      Fig.~\ref{fig:m_ej}. The solid lines correspond to the case where
      the jet is collimated and show the position of the jet's head
      ($z_h$), the radius of the ejecta front ($R_{\rm ej}$), the
      position where the first recollimation shock converges ($z_{cj}$),
      and the cylindrical radii of the jet's head ($r_j$) and the cocoon
      ($r_c$).  The dashed lines show the case when the jet is forced to
      remain uncollimated.}
    \label{fig:jet-cocoon}
\end{figure}
%%%%%%%%%%%%%%%%%%%%%%%%%%%%%%%%%%%%%%%%%

In Fig.~\ref{fig:jet-cocoon}, we show the dynamics of the jet-cocoon
system expanding inside the quasi-isotropic ejecta from the HMNS with
$\beta_{\max}=0.3$, where the relativistic jet is injected at $t=t_{\rm
  coll}=1\,{\rm s}$ with bulk Lorentz factor $\Gamma_j = 10$,
corresponding to an initial half-opening angle of
$\theta_0=\Gamma_j^{-1}\simeq 5.7^\circ$. Here, and in the rest of the
paper, we assume a reference value for the single-sided jet power of
$L_j=E_j/t_{90}=10^{50}~{\rm erg~s}^{-1}$, where $E_j$ is the
single-sided true jet energy and $t_{90}$ is the duration of the prompt
emission within which $90\%$ of the fluence is accumulated. In the case
of GRB~170817A, the prompt gamma-ray emission lasted for
$t_{90}=2.0\pm0.5\,{\rm s}$ \citep{Abbott+17b,Goldstein+17}, where the
main hard spike had a duration of $\sim0.5\,{\rm s}$ and was followed by
a tail of softer emission. Many numerical simulations
\citep[\eg][]{Granot+18,Lazzati+18,Mooley+18b,Xie+18} of this event find
a true jet energy of $E_j \sim 10^{50}\,$ erg that motivated the assumed
value of the jet power in this work.

Immediately after injection, the velocity of the jet's head slows down
significantly to $\beta_h\ll1$ as it encounters the ejected mass. The
pressure in the surrounding cocoon starts to build up and the jet begins
to be collimated, as seen from the reduction in the cylindrical radius
$r_j$ of the jet (see Fig.~\ref{fig:jet-cocoon} for $t-t_{\rm coll}
\lesssim 10^{-4}\,{\rm s}$). This also aids in accelerating the jet due
to the shrinkage in the area of the working surface at the jet's
head. This phase, however, only lasts for a very short time and as the
cocoon expands, shown by the increase in $r_c$, the drop in its pressure
is insufficient to maintain collimation of the jet. At this point, the
size of the jet's head starts to expand and the jet essentially
propagates uncollimated. Therefore, the jet only encounters a single
recollimation shock, and that is only due to the inherent simplicity of
the underlying model. In principle, the jet may experience multiple
recollimation shocks before breakout which would cause $r_j$ to
oscillate. Nevertheless, the half-opening angle of the jet is now smaller
than its initial value, $\theta_j<\theta_0$, and the jet maintains it
until it clears the ejecta and breaks out when $z_h(t)=R_{\rm ej}(t)$ at
a breakout time $t_{\rm br}\approx0.65\,{\rm s}$ in the case considered
in Fig.~\ref{fig:jet-cocoon}. After the first recollimation shock, we
find that the jet head maintains an approximately constant velocity, with
$\beta_h\simeq0.7$, and constant $\tilde L\simeq 6 < \theta_0^{-4/3}$,
until just before breakout, after which point the jet head becomes
relativistic with $\beta_h\simeq1$.

Also shown with dashed lines in Fig.~\ref{fig:jet-cocoon} is the
corresponding dynamics in the case in which the jet is forced to remain
uncollimated, \ie it maintains the same opening angle as the initial one,
and in this case it is clear that the jet takes longer to breakout (by a
factor of $\simeq2$ in time) and might even be choked if the engine shuts off
before that time. Since the solid angle is fixed in this case, the radius
of the working surface grows with the vertical distance traveled by the
jet, \ie $r_j\propto z_h$ at all times.

\begin{figure}
    \centering
    \includegraphics[width=0.48\textwidth]{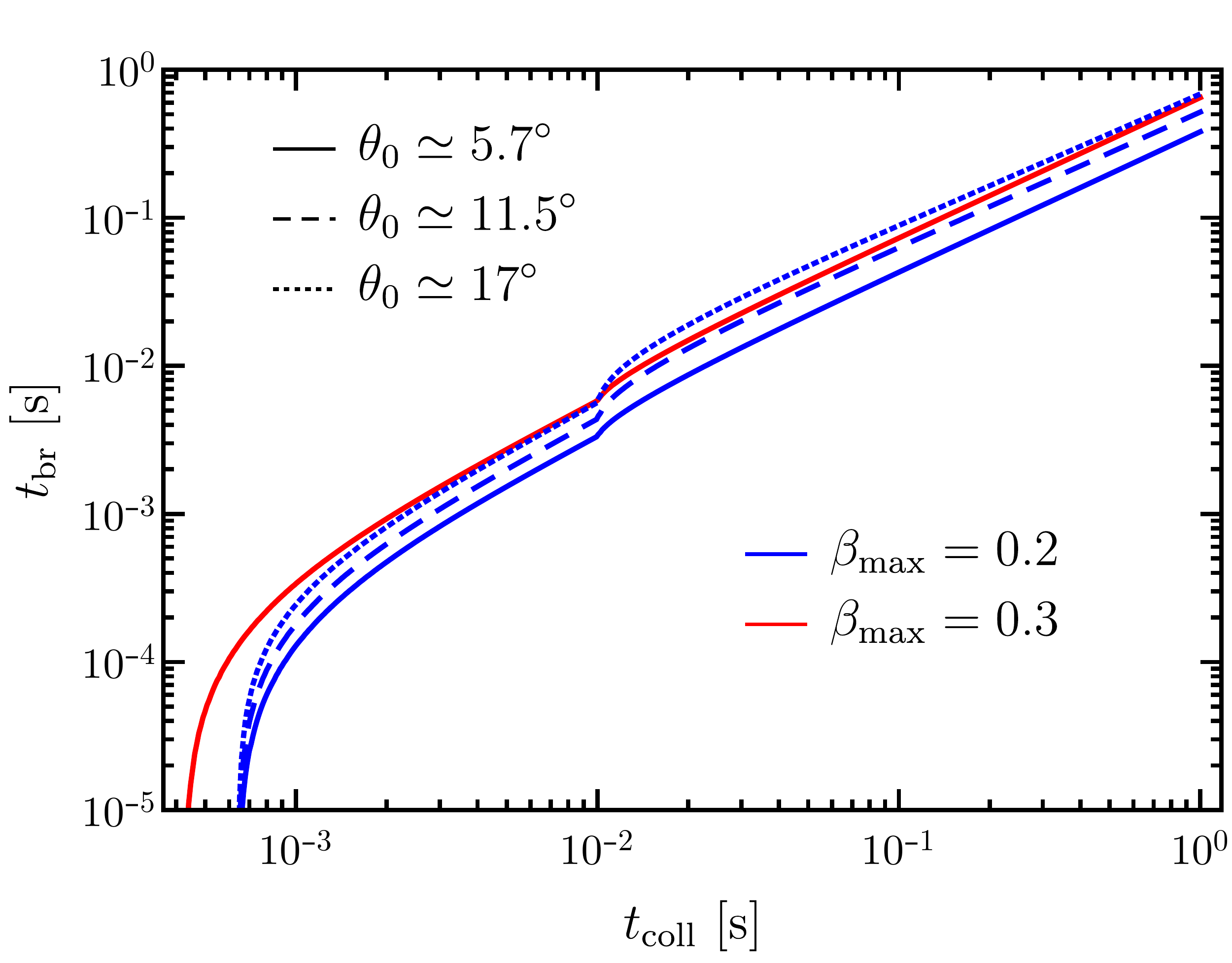}
    \caption{Jet-breakout time $t_{\rm br}$ as a function of the collapse
      time $t_{\rm coll}$ of the HMNS. The rest of the setup is the same
      as in Fig.~\ref{fig:jet-cocoon}. Shown here are the breakout times
      for different values of the initial jet half-opening angle
      $\theta_0=\Gamma_0^{-1}$ and different fastest ejecta velocities
      $\beta_{\max}$. Curves for different values of $\theta_0$ refer to
      $\beta_{\rm max}=0.2$ and the small jump at $t_{\rm coll}=10^{-2}\,{\rm s}$ 
      is due to the enhanced rate of mass ejection by the newly formed HMNS.}
    \label{fig:tbo-tcoll}
\end{figure}

Figure~\ref{fig:tbo-tcoll} shows the jet-breakout time $t_{\rm br}$ as a
function of the collapse time and how the $t_{\rm br}$ depends on the
different ejecta velocities $\beta_{\max}$ and on different jet initial
half-opening angles. The mass-ejection rates are those discussed in
Sec.~\ref{sec:mass-ejection}, so that a longer $t_{\rm coll}$ results in
a larger ejected mass in the path of the jet and thus to a delayed jet
breakout. The sharp drop in $t_{\rm br}$ at early times simply shows that
the ejecta has not had enough time to expand and the jet breaks out
almost immediately. Note that the rate at which the ejecta is expanding
has a significant effect on the breakout time (both scales are
logarithmic), with longer $t_{\rm br}$ in the faster expanding ejecta. A
similar effect can also be produced if the jet has a larger initial
half-opening angle $\theta_0$, which increases the area of the working
surface at the jet's head and causes the jet to slow down. The small jump
at $t_{\rm coll}=10^{-2}\,{\rm s}$ is due to the enhanced rate of mass
ejection by the newly formed HMNS at $t=t_{\rm dyn}\sim10\,{\rm ms}$ post
merger (see left panels of Fig.~\ref{fig:m_ej}).

After breakout, the jet very rapidly accelerates and its energetic core
expands with ultra-relativistic speeds, $\Gamma_j\gg1$, where it is
surrounded by less energetic and relatively slower-moving material. This
phase of the jet dynamics is outside the scope of this work and is not
calculated by the one-dimensional model used above. In the next section,
and for simplicity, the jet is treated as an expanding conical flow, so
that it is convenient to use the radial coordinate $r$ in place of the
cylindrical coordinate $z$.

%%%%%% FIGURE  %%%%%%%%%%%%%%%%%%%%%%%%%%
%%%%%%%%%%%%%%%%%%%%%%%%%%%%%%%%%%%%%%%%%%

%%%%%% CONSTRAINTS ON HMNS LIFETIME AND EOS %%%%%%%%%%%%%%%%%%%%%%%%%%%%%%%%%%%%
\section{Collapse time from EM delay}
\label{sec:constraint-time}

The delay of $1.74\,{\rm s}$ in the onset of the prompt emission measured
in GRB~170817A can be explained by a sum of three different timescales
\citep[\eg][]{Granot+17}. First, and most obviously, the launching of the
jet is delayed by the HMNS that will survive for a timescale $t_{\rm
  coll}$ (see Sec. \ref{subs:kilonova}), after which it will collapse to
a BH. Second, the jet has to clear the ejecta before it can radiate the
prompt emission photons, otherwise the Thomson optical depth of the
medium would be too large and it would suppress any high-energy
emission. The jet clears the ejecta over the breakout time $t_{\rm br}$
that, in turn, depends on $t_{\rm coll}$, as well as the total ejected
mass, density profile, and the jet-launching properties. Third, since GWs
propagate at the speed of light \citep{Abbott+17b} and the jet propagates
at slightly lower velocities, there is always a radial time delay $t_R$
between the two components to reach the prompt gamma-ray emission radius
$R_\gamma$, \ie
\begin{equation}
  t_R = (1+z_c) \frac{R_\gamma}{2\Gamma^2c} \simeq \frac{R_\gamma}{2\Gamma^2c}\,,
\end{equation}
where in the second equality we have accounted for the fact that the
considered cosmological redshift $z_c$ of GW170817 is
essentially zero. As a result of these considerations, the total delay
time can be written as
\begin{equation}
  \label{eq:tdelay}
t_{\rm del} = 1.74 \pm 0.05\,{\rm s} = t_{\rm coll} + t_{\rm br}(t_{\rm coll}) +
t_R\,,
\end{equation}
where we have expressed explicitly that the jet-breakout time depends on
the collapse time of the remnant. The breakout time considered above is
that for jet material that breaks out of the ejecta along the jet
symmetry axis. In our model, we assume that after breakout the
relativistic jet core is surrounded by a sheath of lower energy and
slower moving material \citep[e.g.,][]{Kathirgamaraju+18} which undergoes
dissipation. Alternatively, some works
\citep[e.g.,][]{Kasliwal+17,Gottlieb+17,Nakar+18} have argued that the
prompt gamma-ray emission was produced by the breakout of the shock
driven by the expanding cocoon from the sub-relativistic circum-merger
ejecta. In this case, the breakout time at an angle $\theta_{\rm obs}$
away from the jet symmetry axis will be longer and consequently the
collapse time shorter.

At $r=R_\gamma$ the jet experiences internal dissipation and produces the
prompt gamma-ray emission. The underlying energy dissipation mechanism is
still uncertain \citep[see \eg][for a review]{Kumar-Zhang-15}.  If the
jet is kinetic-energy dominated, energy can be dissipated in internal
collisions of mass shells ejected by the BH with variable Lorentz
factors, where faster shells coming from the engine would catch up with
slower shells at the dissipation radius. Alternatively, if the jet is
Poynting-flux dominated, magnetic energy is the main energy reservoir,
which can be tapped via magnetic reconnection or MHD instabilities.

The condition for causality dictates that
$R_\gamma\lesssim2\Gamma^2c\delta t_{\rm obs}$, where $\delta t_{\rm
  obs}$ is the observed flux-variability time
\citep[\eg][]{Sari-Piran-97}. The same condition is also obtained in the
internal-collision scenario, in which case $\Gamma$ is the Lorentz factor
of the slower-moving shell and $\delta t_{\rm obs}$ represents the
engine-variability time. The observed variability time of the GRB
lightcurve can be associated to either the rise time of the pulse, or the
decay time, or their combination, depending on their contribution to the
total pulse width, which can be expressed as $\Delta t_{\rm pulse} :=
\delta t_r + \delta t_\theta$. To be more specific, consider a thin,
spherical, relativistic shell that emits over a range of radii
$(R_\gamma-\delta R)\leq r\leq R_\gamma$. The emission from this shell
will be temporally spread out over the timescale $\delta t_r=\delta
R/2\Gamma^2c$ and will represent the rise time of the pulse. On the other
hand, the decay time of the pulse occurs due to light travel-time effects
across a curved emitting surface, also referred to as the ``angular
time'', such that $\delta t_\theta = R_\gamma(1-\cos\tilde\theta)/c$,
where the term in the parentheses represents the extra distance traveled
by the photon that was emitted at angle $\tilde\theta$ away from the 
line-of-sight. For a relativistic shell, the emission is beamed within a cone
of angle $\tilde\theta\simeq\Gamma^{-1}\ll1$, which yields $\delta
t_\theta\simeq R_\gamma/2\Gamma^2c$. This finally yields
\citep[\eg][]{Genet-Granot-09},
\begin{equation}
\Delta t_{\rm pulse} = \frac{R_\gamma}{2\Gamma^2c}\left(1+\frac{\delta
  R}{R_\gamma}\right)\,.
\end{equation}

The minimum variability time for the prompt gamma-ray emission in
GRB~170817A, which would correspond to the rise time to the pulse peak,
\ie $\delta t_r$, measured by Fermi/GBM in the energy range of
$50-300\,{\rm keV}$ is $\delta t_{\min}=0.125\pm0.064\,{\rm s}$ and the
pulse decay time is $\delta t_\theta\approx0.5\,{\rm s}$
\citep{Abbott+17b,Goldstein+17}.  On the other hand, INTEGRAL/SPI-ACS
observed the main hard spike spread out over two time bins of $50\,{\rm
  ms}$, with a total duration of $\Delta t_{\rm pulse}\lesssim100\,{\rm
  ms}$ in the $\sim75-2000\,{\rm keV}$ energy range
\citep{Savchenko2017}. These two measurements suggest that the pulse
duration is energy dependent and that the decay time of the pulse $\delta
t_\theta\lesssim100\,{\rm ms}$ at high energies. Unless $\delta
R/R_\gamma\approx1$ the pulse rise time cannot be used to robustly
determine the emission radius. Therefore, we associate the variability
time to the angular time, $\delta t_{\rm obs}\approx \delta t_\theta$,
and use it to constrain the emission radius given an estimate of
$\Gamma$.

Almost all distant GRBs are observed within the semi-aperture of the
bright core, such that $\theta_{\rm obs}\lesssim\theta_c$, where
$\theta_{\rm obs}$ is the viewing angle and $\theta_c$ is the
half-opening angle of the bright core (both angles are measured from the
jet symmetry axis). In case the structure of the jet can be modeled as
having sharp edges, then $\theta_c=\theta_j$, where $\theta_j$ is the
half-opening angle of the jet. In these GRBs, compactness arguments
\citep[\eg][]{Lithwick-Sari-01,Gill-Granot-18a} dictate that
$\Gamma>100$, otherwise opacity due to $\gamma\gamma$-annihilation (\ie
$\gamma\gamma\to e^-+e^+$) would prevent the radiation of gamma-ray
photons. Modeling of the radio/optical/X-rays afterglow emission from
GRB~170817A, both using semi-analytical models
\citep[\eg][]{Gill-Granot-18b,Lamb-Kobayashi-18,Troja+18}, as well as
numerical simulations \citep[\eg][]{Granot+18,Lazzati+18,Margutti+18,Kathirgamaraju+19} 
of structured jets, has revealed that the line-of-sight is significantly
off-axis, with $20^\circ\lesssim\theta_{\rm obs}\lesssim28^\circ$ and
$\theta_c\sim5^\circ$.

The prompt gamma-ray emission of GRB~170817A at a distance of $40\,{\rm
  Mpc}$, $D_{40}$, with isotropic-equivalent energy of $E_{\gamma,\rm
  iso}=(5.36\pm0.38)\times10^{46}D_{40}^2\,$erg, was $\sim3$-$4$ orders
of magnitude lower than what is typically observed from distant
short-hard GRBs. This fact alone had already hinted at an off-axis
jet. Early analysis of the burst energetics showed that the prompt
gamma-ray emission was dominated by low-energy material along the
line-of-sight \citep[\eg][]{Granot+17} and with only modest $\Gamma$. By
applying the compactness arguments, it was further realized that the
emission region must be moving towards the observer with $\Gamma\gtrsim5$
\citep{Matsumoto+18}. This lower limit on $\Gamma$ can now be used to
obtain a constraint on the dissipation radius \citep[also see][]{Beloborodov+18},
\begin{equation}
    R_\gamma \gtrsim 7.5\times10^{11}\pfrac{\Gamma}{5}^2
    \pfrac{\delta t_{\rm obs}}{0.5\,{\rm s}}\,{\rm cm}\,.
\end{equation}
If the decay time of the pulse is indeed as short as $\sim100\,{\rm ms}$
then the dissipation radius can be smaller by a factor of $\sim5$.

%%%%%%%% FIGURE  %%%%%%%%%%%%%%%%%%%%%
\begin{figure}
    \centering
    \includegraphics[width=0.48\textwidth]{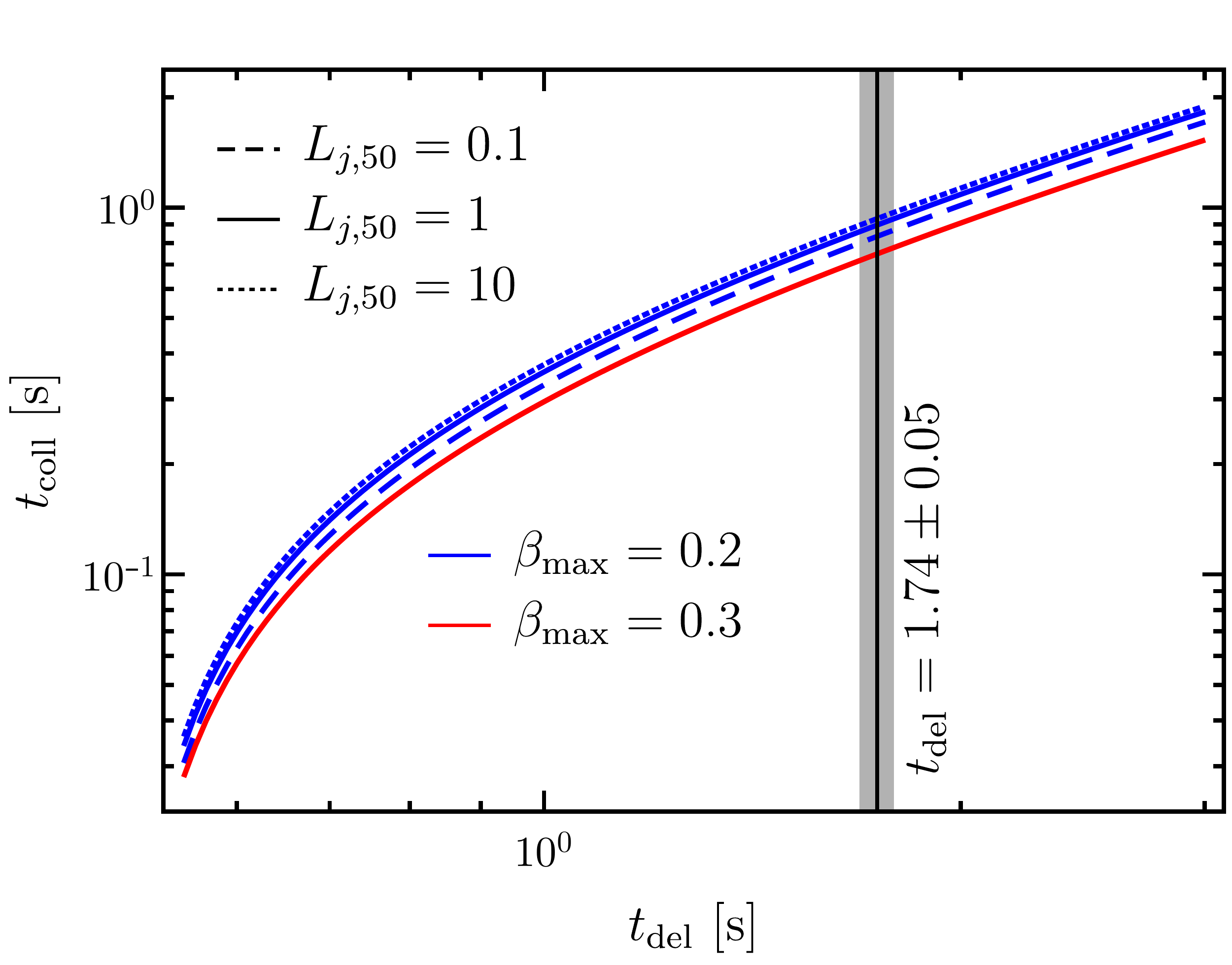}
    \caption{The collapse time as a function of the delay time between
      the GW chirp signal and the onset of the prompt gamma-ray
      emission. Different lines refer to different $\beta_{\max}$ and jet
      powers, for $k=2$ and $\theta_0\simeq 5.7^\circ$. The thin vertical
      line marks $t_{\rm del}=1.74\,{\rm s}$ that was observed for
      GRB~170817A, and it shows a collapse time of $t_{\rm
        coll}\simeq0.75\,{\rm s}$ obtained for $\beta_{\max}=0.3$ and
      $L_j=10^{50}\,{\rm erg~s}^{-1}$.}
    \label{fig:tcoll-plt}
\end{figure}
%%%%%%%%%%%%%%%%%%%%%%%%%%%%%%%%%%%%%%

The angular time also yields the radial delay time $t_R=\delta t_\theta$
between the relativistic ejecta and the GWs to arrive at the emission
radius $R_\gamma$. From Eq.~(\ref{eq:tdelay}), we find that a longer
$t_R$ would yield a shorter collapse time. Therefore, to obtain a
conservative estimate of $t_{\rm coll}$, we use $t_R\approx0.5\,{\rm
  s}$. Using this, we next numerically solve Eq.~(\ref{eq:tdelay}) for a
given jet luminosity and mass-ejection rate to obtain the collapse time
of the HMNS as a function of $t_{\rm del}$. This is shown in
Fig.~\ref{fig:tcoll-plt}, where the different lines refer to different
values of $\beta_{\max}$ and of the jet power, using as reference values
$k=2$ and $\theta_0\simeq 5.7^\circ$. The thin vertical line marks
$t_{\rm del}=1.74\,{\rm s}$ that was observed for GRB~170817A, and it
shows a collapse time of $t_{\rm coll}\simeq0.75\,{\rm s}$ obtained for
$\beta_{\max}=0.3$ and $L_j=10^{50}\,{\rm erg~s}^{-1}$. In this way, we
deduce that the collapse time as constrained from the EM delay must have
been at
\begin{equation}
\label{eq:tcoll_EM}
  t_{\rm coll}=0.82\pm 0.15\ \,{\rm s}\,,
\end{equation}
where the range in the variance is dictated from considering $\beta_{\rm
  max}=0.2$ or $\beta_{\rm max}=0.3$, respectively. This result is robust
in regards to the jet power and changes only weakly as the jet power is
varied by an order of magnitude. Furthermore, varying the density profile
by using different values of the power-law index $k$ in
Eq.~\eqref{eq:rho_ej} yields only small differences.

%%%%%%%% FIGURE  %%%%%%%%%%%%%%%%%%%%%
\begin{figure}
    \centering
    \includegraphics[width=0.48\textwidth]{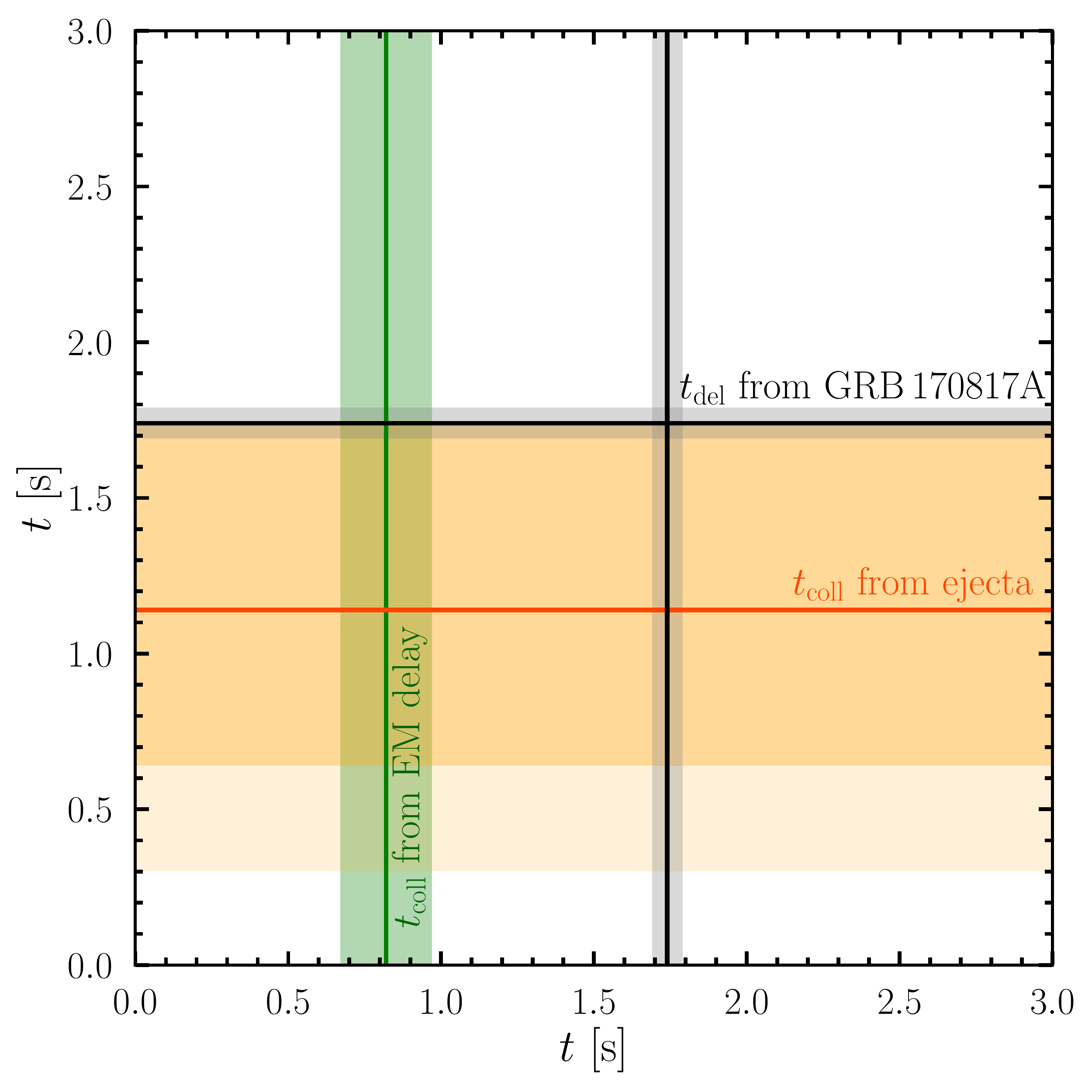}
    \caption{Comparison of the HMNS collapse times obtained from two
      distinct lines of argument. The horizontal orange line and the
      vertical green line report the estimated times for the collapse of
      the remnant in GW170817 as deduced from the time needed for the jet
      to make its way across the ejected matter. In both cases, the
      corresponding uncertainty regions are shown with shaded regions of
      the corresponding colour; the light orange-shaded area refers to
      the more conservative lower limit on $t_{\rm coll}$ when
      considering an uncertainty in the power-law exponent in the
      mass-ejection rate. Also shown is the delay time (black solid line)
      and the corresponding uncertainty (black shaded area).}
    \label{fig:tcoll-compare-plt}
\end{figure}
%%%%%%%%%%%%%%%%%%%%%%%%%%%%%%%%%%%%%%

%%%%%%% DISCUSSION %%%%%%%%%%%%%%%%%%%%
\section{Discussion}
\label{sec:discussion}

Figure~\ref{fig:tcoll-compare-plt} offers a synthetic and yet
comprehensive summary of the results of this paper by comparing the
estimates on the collapse time of the remnant of GW170817. More
specifically, the horizontal orange line reports the estimated collapse
time, $t_{\rm coll}$, as deduced from the ejected matter producing the
blue-kilonova emission [\cf Eq.~\eqref{eq:tcoll_ejecta}]. Similarly, the
vertical green line reports the estimate on $t_{\rm coll}$ needed for the
jet to burrow its way through the intervening ejecta [\cf
  Eq.~\eqref{eq:tcoll_EM}]. In both cases, the corresponding uncertainty
regions are shown with shaded regions of the corresponding colour; the
light orange-shaded area refers to the more conservative lower limit on
the collapse time obtained when considering an uncertainty in the
power-law exponent in the mass-ejection rate (see
Sec. \ref{sec:tcoll-ejecta}). Also shown is the delay time between the GW
and EM emission, $t_{\rm del}$, (black solid line) and the corresponding
uncertainty (black shaded area). Combining this information, we deduce
that the best estimate for the collapse time of the remnant of GW170817
is
%\footnote{A more conservative lower limit on the collapse time is
%  obtained, with $t_{\rm coll}=0.98^{+0.31}_{-0.43}\ \,{\rm s}\,$, when
%  we consider a $30\%$ uncertainty in the power-law exponent in the
%  mass-ejection rate (\ref{m_B_HMNS}) from magnetically driven winds.}
%
\begin{equation}
\label{eq:tcoll_combined}
  t_{\rm coll}=0.98^{+0.31}_{-0.26}\ \,{\rm s}\,.
\end{equation}

A more conservative lower limit on the collapse time is obtained, with
$t_{\rm coll}=0.98^{+0.31}_{-0.43}\ \,{\rm s}\,$, when we consider a
$30\%$ uncertainty in the power-law exponent in the mass-ejection rate
(\ref{m_B_HMNS}) from MHD-driven winds. The level of uncertainty
considered here in the power-law exponent is more conservative than that
inferred from simulations, which is $\approx11\%$ and for which we find a
less conservative lower limit on the collapse time as compared to that
obtained from a constant fractional error on the ejected mass. Due to the
small number of simulations that have explored the MHD-driven mass
ejection channel, the magnitude of the uncertainty in the power-law
exponent is poorly known at best. Therefore, we adopt as a reference the
estimate in Eq. (\ref{eq:tcoll_combined}) for the collapse time in this
work.

Such a survival time for the remnant is considerably longer than the one
explored in three-dimensional general-relativistic simulations of BNS
mergers, which so far have been limited to simulations with $t_{\rm
  coll}\sim10$-$100\,{\rm ms}$. Since on timescales $t_{\rm coll}
\lesssim 100\,{\rm ms}$ the most efficient process to remove energy and
angular momentum is the emission of GWs, the implication stemming from
the estimate \eqref{eq:tcoll_combined} is that other, less-efficient
mechanisms, need to be invoked to extract angular momentum on such long
timescales. At the same time, magnetic spin-down provides a simple and
compelling explanation. Using Eq.~(\ref{eq:t_sd}) for a remnant mass of
$M \approx 2.7\,M_\odot$, it is simple to deduce that a spin-down
timescale of the order of the collapse time, \ie $t_{\rm sd} \sim t_{\rm
  coll} \sim 1\,{\rm s}$, can be accomplished with magnetar-strength
surface magnetic fields, $B_s \simeq 10^{16}\,{\rm G}$. In addition, such
surface magnetic fields would naturally provide an enhanced rate of mass
loss in the polar region due to magnetocentrifugal emission
\citep{Metzger2018} and a high rate of expansion of the ejecta, with $0.2
\lesssim \beta_{\max} \lesssim0.3$. Interestingly, the collapse of the
remnant when such large magnetic fields are present would also lead to an
explosive outflow \citep{Salafia2017, Nathanail2018} (see also
\citealt{Nathanail2018c} for a discussion of the observable signatures).

At the same time, magnetic fields in the post-merger object cannot reach
the largest values allowed by energy equipartition, \ie $B_{\rm int}
\simeq 10^{17}$-$10^{18}\,{\rm G}$. Such ultra-strong magnetic fields, in
fact would prolong the breakout time as the jet would have to plow
through more material and spend more time catching up with the expanding
ejecta front. Furthermore, the duration distribution of short-hard GRBs
are suggestive of a characteristic jet breakout time of $t_{\rm br}
\simeq 0.4\,{\rm s}$ \citep{Moharana-Piran-17}; assuming that this
timescale is indeed common to all short GRBs, this would limit the
collapse time to $t_{\rm coll} \sim 0.7\,{\rm s}$ for a typical jet power
of $L_j = 10^{50}\,{\rm erg~s}^{-1}$. The total time delay in this case
would be of the order of $t_{\rm del}\gtrsim1\,{\rm s}$, which can be
confirmed by future multi-messenger detections of BNS mergers. In
summary, strong (but not ultra-strong) magnetic fields represent a
natural explanation for the high mass-ejection rates and the long delay
times between the GW and the gamma-ray emission.

Although the production of such magnetic fields still represents a
serious challenge for direct numerical simulations, the existence of a
long-lived HMNS can in principle also be confirmed by the detection of
GWs after the merger. The GW signal, in fact, has clear spectroscopic
features, with precise oscillation frequencies in the range
$\sim2-4\,{\rm kHz}$, and which can be related to the EOS and the
properties of the progenitors \citep[e.g.,][]{Bauswein2011, Takami2014,
  Rezzolla2016}. As mentioned earlier, in the case of GW170817, the
search for GWs from the remnant in the signals from both advanced LIGO
and Virgo detectors came out empty \citep{Abbott+17d} (but see
\citealt{vanPutten2019} for a different claim). Future detections of BNS
mergers by advanced or third-generation detectors such as the Einstein
Telescope \citep{Punturo2010b,Sathyaprakash+12}, will provide better
estimates of the collapse time of the remnant and help clarify what is
the dominant process of removal of angular momentum on such long
timescales.

%%%%%%% CONCLUSIONS %%%%%%%%%%%%%%%%%%%
\section{Conclusions}
\label{sec:conclusions}

The GW event GW170817 and its electromagnetic counterpart, GRB~170817A,
have not only marked the birth of multi-messenger GW astronomy, but have
also confirmed that mergers of binary systems of neutron stars are
responsible for the phenomenology of short GRBs. Unfortunately, because
of its high-frequency properties during the final stages of the inspiral,
the GW signal from GW170817 became invisible to the LIGO/Virgo detectors
before the actual merger took place. As a result, we are unable to
extract direct GW information about the object that was produced by the
merger and that, most likely, was a metastable HMNS.

Making use of the observational features of GRB 170817A, namely, that the
main hard pulse of gamma-ray emission had a duration of $\sim0.5\,{\rm
  s}$ and its onset was delayed by $t_{\rm del} \approx 1.74\,{\rm s}$
with respect to the GW chirp signal, and assuming that the prompt
gamma-ray emission is ultimately the result of a relativistic jet powered
by a rotating BH, we have constrained the lifetime of the remnant of
GW170817. In particular, we have used the properties of the kilonova
emission and the delay time between the GW chirp signal and prompt
gamma-ray emission onset to constrain the collapse time of the HMNS. We
have argued that in order to produce the requisite mass of $M_{\rm
  ej,blue} \approx 0.025\,M_\odot$ in the lanthanide-poor ejecta that
gave rise to the rapidly fading bluer emission in the UV and optical at
early times, the collapse time of the HMNS cannot be different from about
one second. More importantly, we have reached a very similar conclusion
from an independent line of argument, where we model the dynamical
evolution of the relativistic jet launched after the HMNS collapses and
bores its way out of the dynamical ejecta.

To derive the estimates on the collapse time from the ejected matter, we
have combined into a single systematic description the various processes
leading to the ejection of matter from the remnant, and that include: the
dynamical ejecta, the neutrino, as well as the MHD-driven winds from both
the HMNS and the accretion torus. Estimates for the mass-ejection rates
from all of these channels have been extracted from numerical-relativity
simulations carried out on much shorter timescales and which have been
extrapolated to the timescales of interest after casting them into
suitable analytic forms. In this way, we have deduced that the collapse
time is constrained to be $t_{\rm coll}=1.14^{+0.60}_{-0.50}\ \,{\rm
  s}$. Similarly, to derive the collapse time starting from the delay
between the GW chirp signal and the onset of the prompt gamma-ray
emission, we have computed the jet-breakout time for a fiducial set of
parameters for the homologously expanding ejecta, \eg an expansion
velocity $0.2\leq\beta_{\max}\leq0.3$ and a jet luminosity $0.1\leq
L_{j,50}\leq10$. These estimates are less uncertain than those related to
the ejection of mass and have allowed us to deduce that the collapse time
is constrained to be $t_{\rm coll}=0.82\pm 0.15\ \,{\rm s}$.  Combining
these two results and the corresponding uncertainties, we conclude that
the remnant of GW170817 must have collapsed to a rotating BH at about
$t_{\rm coll} = 0.98_{-0.26}^{+0.31}\ \,{\rm s}$ after
merger. Interestingly, this estimate represents the first attempt to use
GW and astronomical observations to constrain the properties of the
gravitational collapse to a BH.

Looking at the future, the collection of new multi-messenger GW
detections in the coming months will provide additional important
information on the delay between the GW and the EM signals and hence
better constrain the collapse time of the HMNS for a given mass of the
binary system. In turn, this will bring clarity to many open questions
not only on the merger of BNSs, but also on the mechanism leading to the
production of a short GRB. First, they will help to narrow down the
search for the correct EOS, since the collapse time depends on it, with
some EOSs yielding only a short-lived HMNS (\ie $t_{\rm
  coll}\lesssim10\,{\rm ms}$), while others producing a longer-lived
remnant. Second, it will provide a better handle on the dominant
dissipative processes, \eg energy and angular momentum loss due to GW
radiation dominates for $t_{\rm coll}\lesssim100\,{\rm ms}$, whereas
magnetic braking due to strong internal fields dominates
thereafter. Third, since the large majority of the ejected matter leading
to the blue-kilonova signal is expelled before the remnant collapses to a
BH, a clear correlation between $t_{\rm coll}$ and $M_{\rm ej,blue}$ is
expected and can be used to constrain the properties of the remnant.
Finally, if a long-lived HMNS is produced, then the collapse time will
yield a constraint on the internal magnetic field and provide further
insight into the physics of the dissipative effects that drive the HMNS
to uniform rotation. In future work we plan to explore several of these
issues, both through self-consistent numerical simulations and via
semi-analytical modelling.

\acknowledgments
It is a pleasure to thank Jonathan Granot, Elias Most, and David Radice for
useful discussions and input. Support comes from: the ERC synergy grant
``BlackHoleCam: Imaging the Event Horizon of Black Holes'' (Grant
No. 610058), ``PHAROS'', COST Action CA16214; the LOEWE-Program in HIC
for FAIR; the European Union's Horizon 2020 Research and Innovation
Programme (Grant 671698) (call FETHPC-1-2014, project ExaHyPE).

\bibliographystyle{yahapj}
\bibliography{refs,aeireferences} % if your bibtex file is called example.bib

\end{document}